\documentclass[aps,pre,superscriptaddress]{revtex4}
\usepackage[dvips]{graphics}
\usepackage{graphicx}
\usepackage{amsfonts}
\usepackage{amssymb}
\usepackage{amsmath}
\usepackage{subfigure}
\usepackage{color}

\begin{document}

\title{Collisional-inhomogeneity-induced generation of matter-wave dark solitons}

\author{C. Wang}
\affiliation{
Department of Mathematics and Statistics, University of Massachusetts, Amherst, MA 01003-4515}
\author{P. G. Kevrekidis}
\affiliation{
Department of Mathematics and Statistics, University of Massachusetts, Amherst, MA 01003-4515}
\author{T. P. Horikis}
\affiliation{
 Department of Mathematics, University of Ioannina, Ioannina 45110, Greece}
\author{D. J. Frantzeskakis}
\affiliation{
Department of Physics, University of Athens, Panepistimiopolis, Zografos, Athens 15784, Greece}

\begin{abstract}
We propose an experimentally relevant protocol for the controlled generation of matter-wave dark solitons
in atomic Bose-Einstein condensates (BECs). In particular, using direct 
numerical simulations, we show that by switching-on a spatially inhomogeneous (step-like) change of the $s$-wave scattering length, it is possible to generate a controllable number of dark solitons in a quasi-one-dimensional BEC. A similar phenomenology is also found in the two-dimensional setting of ``disk-shaped'' BECs but, as the solitons
are subject to the snaking instability, they decay into vortex structures. A detailed investigation of how the parameters involved affect the emergence and evolution of solitons and vortices is provided.
\end{abstract}

\maketitle

\section{Introduction}

The experimental realization of Bose-Einstein condensates (BECs) of dilute atomic gases \cite{BEC}
is one of the most fundamental developments in quantum and atomic physics over the last two decades.
Many of the relevant theoretical and experimental studies \cite{BECBook,book2} have been focused on
the nonlinear excitations of BECs \cite{revnonlin,ourbook}. Among them, the so-called
{\it matter-wave dark solitons} (see the recent review \cite{djf}), namely localized density dips with a phase-jump across them, have been observed in BECs in a series of experiments \cite{han1,nist,dutton,bpa,ginsberg2005,engels,hamburg,hambcol,kip,technion,draft6}.
The interest on this type of localized nonlinear structures may be attributed to a series of reasons:
matter-wave dark solitons are one of the most fundamental nonlinear 
excitations of BECs with repulsive
interatomic interactions, being the nonlinear analogs of the excited
eigenstates of the quantum harmonic oscillator \cite{KivsharPLA,JPB}; 
furthermore, 
they arise spontaneously upon crossing the BEC phase-transition \cite{zurek2,zurek3}, and their properties may be used as a diagnostic tool probing the rich physics of BECs \cite{anglin}. Furthermore, matter-wave dark solitons may also be relevant to applications: for example, it has been proposed that the position of a dark soliton can be used to monitor the phase acquired in an atomic matter-wave interferometer in the nonlinear regime \cite{appl1,appl2}.

Matter-wave dark solitons have been generated in atomic BECs by means of various different methods. In particular, in the early experiments \cite{han1,nist} (but also in some recent ones \cite{hamburg,hambcol}), dark solitons were created by means of the {\it phase-imprinting method}: this technique relies on the illumination of part of the condensate by a short off-resonance laser beam, so as to imprint the necessary phase jump characterizing the dark soliton. Another technique, that was used in the experiments reported in Refs.~\cite{dutton,ginsberg2005} (see also \cite{enghoef}) is the {\it density-engineering method}, which is based on the creation of local density depletion on the BEC density by means of external potentials. A combination of the phase-imprinting and density-engineering methods, namely the so-called {\it quantum-state engineering} technique is also possible: this method involves manipulation of both the BEC density and phase, and has also been used in experiments with two-component BECs \cite{bpa,hamburg}. Furthermore, in Ref.~\cite{engels}, the BEC flow against a broad penetrable barrier, which was swept through the condensate, led to the breakup of the BEC superfluidity and the concomitant generation of dark solitons for a particular regime of barrier speeds. Finally, in more recent experiments the the interference between two (or more) condensate fragments confined in a trap was demonstrated to result in the formation of dark solitons \cite{kip,technion,draft6,enghoef}, or vortices in higher-dimensional setups \cite{brian,nate}.

In this work, we propose an alternative method to create dark solitons in quasi one-dimensional (1D) BECs, or vortices in higher-dimensional settings 
(particularly, in quasi two-dimensional (2D) BECs). The proposed technique relies on the manipulation of the $s$-wave scattering length, which controls the effective nonlinear coefficient in the Gross-Pitaevskii (GP) mean-field model: in particular, we consider
a situation where the collisional dynamics across the condensate is inhomogeneous due to the presence of a spatially dependent external field in the vicinity of a Feshbach resonance. Note that such, so-called, {\it collisionally inhomogeneous condensates} \cite{gofx}, have attracted much attention, as they have been proved to provide a variety of interesting phenomena \cite{gofxo1,gofxo2,gofxo3,gofxo4,gofxo5,gofxo6,gofxo7,gofxo8,gofxo9,gofxo10,gofxo11,gofxo12,gofxo13}.
In the present study, we assume, more specifically, that by switching-on an external magnetic or optical field, the nonlinear coefficient becomes piecewise constant (in line with the considerations of \cite{gofxo2,gofxo3,gofxo12,gofxo13}), and can vary between two (smoothly connected) values. Then, depending on the magnitude of the jump in the value of the nonlinear coefficient, one or more dark solitons in the 1D setting, or vortices in the 2D setting, can controllably be created. It is important to note that the considered form of the spatial change of the nonlinear coefficient (i.e., of the $s$-wave scattering length) may be straightforwardly experimentally realizable, through the use of magnetic field gradients of moderate size for atom chips (see also discussion in Ref.~\cite{gofxo13}).
It should also be added that a similar technique has been proposed
very recently in the three-dimensional context as a means of generating
vortex rings \cite{vict_new}.

The paper is organized as follows. In Sec.~\ref{1D} we present our model and discuss the dark soliton generation in the quasi-1D setting. In Sec.~\ref{2D} we present results concerning the soliton generation, as well as their subsequent decay into vortices, in the quasi-2D setting and, finally, in Sec.~\ref{conc} we present our conclusions.

\section{The model and its consideration in the 1D setting}
\label{1D}
\subsection{Presentation of the model}
We consider a cigar-shaped (quasi-1D) condensate, confined in an anisotropic trap
with longitudinal and transverse confining frequencies (denoted by $\omega_x$ and $\omega_{\perp}$, respectively) such that $\omega_x \ll \omega_{\perp}$. In this case, taking also into regard the quasi-one-dimensional nature of the 
dynamics,
it can be found \cite{gerbier,Delgado} that the condensate evolution can be described by an effectively 1D, GP-like equation which can be expressed in the following dimensionless form:
\begin{equation}
i \partial_{t} u = -\frac{1}{2}\partial_{x}^{2}u + V(x)u + \sqrt{1+g|u|^{2}}u - \mu u.
\label{eq1}
\end{equation}
Here, $u(x,t)$ is the macroscopic wave function, normalized so that $\int_{-\infty}^{+\infty} |u|^2 dx = N$
(where $N$ is the number of atoms), $\mu$ is the chemical potential, while the density $|u|^2$, the length, time and energy are measured, respectively, in units of $(2a)^{-1}$ (where $a$ is the $s$-wave scattering length), the transverse harmonic oscillator length $\alpha_{\perp} \equiv \sqrt{\hbar/m\omega_{\perp}}$, $\omega_{\perp}^{-1}$, and $\hbar \omega_{\perp}$. Note that the deviation from 1D is accounted for by the generalized nonlinearity in Eq.~(\ref{eq1}), which becomes the traditional cubic nonlinearity in the weakly-interacting limit of $g|u|^2 \ll 1$ (in this limit, Eq.~(\ref{eq1}) is reduced to the usual 1D GP equation -- see, e.g., discussion in \cite{revnonlin} and references therein). The potential $V(x)$ in Eq.~(\ref{eq1}) is assumed to have the usual harmonic form,
\begin{equation}
V(x) = \frac{1}{2}\Omega^{2}x^2,
\label{eq2}
\end{equation}
characterized by its strength $\Omega = \omega_x/\omega_{\perp}$. Finally, as concerns the nonlinear coefficient $g$, it is considered
to be of the following form:
\begin{equation}
g \equiv g(x;t) = \Big\{ \frac{1}{2}\varepsilon \Big[1-\tanh\Big(\frac{x}{s}\Big)\Big] \Big\}
\Big\{ \frac{1}{2}\Big[1+\tanh\Big(\frac{t-t_{0}}{\tau}\Big)\Big] \Big\} +1,
\label{eq3}
\end{equation}
where the spatial and temporal scales $s$ and $\tau$, the characteristic time $t_0$, and the constant $\varepsilon$, are defined below, upon considering various asymptotic limits associated to the assumed form of $g(x;t)$. In particular, in order to better understand the physical situation described by Eq.~(\ref{eq3}), let us first consider the asymptotic limit of $t \ll t_0$: in this case, Eq.~(\ref{eq3}) describes a condensate with a constant nonlinear coefficient equal to $g=1$. On the other hand, in the limit $t \gg t_0$, the nonlinearity coefficient asymptotes to the spatially dependent form
$g_{0}(x)=(1/2)\varepsilon[1-\tanh(x/s)]+1$; this suggests, in turn, that for $x \rightarrow \pm \infty$ the
nonlinearity coefficient takes the values $g=1$ and $g=1+\varepsilon$, respectively. In other words,
the condensate is initially ($t \ll t_0$) characterized by a constant nonlinear coefficient $g=1$, but for large times ($t \gg t_0$) the implementation of a temporal smooth ramp renders the nonlinear coefficient spatially inhomogeneous, experiencing a smooth jump, from $g=1+\varepsilon$ to $g=1$, around the trap center ($x=0$). According to the above, it is clear that $\varepsilon$ denotes the amplitude of the jump in the value of the nonlinear coefficient, $s$ is the spatial length scale on which the transition between the values $g=1$ and $g=1+\varepsilon$ takes place, $t_0$ represents the characteristic time around which the spatial inhomogeneity of the nonlinear coefficient is ``switched-on'', while the time scale $\tau$ is the duration
of this process.

From a physical point of view, the above situation may be implemented experimentally as follows.
First, the modification of the nonlinear coefficient $g$ -- i.e., of the scattering length $a$ -- from the value $g=1$ to the value $g=1+\varepsilon$ can be achieved by using an external magnetic (or optical) field, which is 
realized around $t=t_0$. The desired inhomogeneity of the scattering length may be realized upon employing a bias homogeneous external field and imposing on top of it a steep localized spatial gradient. This leads to a constant scattering length in the left portion of the condensate ($x<0$), followed by a localized change of $a$, ending up with a different value in the right portion of the condensate ($x>0$). In our consideration, we choose the function $\tanh$ to mathematically characterize the switch-on of the external fields, as well as the 
transition between the different constant values of $a$.
%, since -- in practice -- there are no ideal steps, but rather close approximations to it. 
Also, it is natural to assume that we are relatively close to a Feshbach resonance, so as to easily manipulate the scattering length with the imposed external field.

\subsection{Dark soliton generation}

Let us now proceed with the presentation of our numerical results demonstrating the dark soliton generation in the 1D setting considered above. The numerical 
simulations start with the determination of the stationary ground state of the system, obtained by means of Newton's method, with an initial condition corresponding to the value $g=1$ of the nonlinear coefficient. Then, the dynamical evolution of the system is monitored by a 4th-order Runge-Kutta method, with step $\Delta t=0.001$. In the results presented below, the parameters are chosen as: $\Omega=0.02$, $t_{0}=10$, $\tau=1$ and $s=1$. Notice that for the generation of dark solitons, the values of the chemical potential and of the jump of the nonlinearity coefficient
(described by the parameters $\varepsilon$ and $\mu$, respectively) are of crucial importance.

\begin{figure}[tbp]
\centering
\includegraphics[width=.85\textwidth,height=.17\textheight]{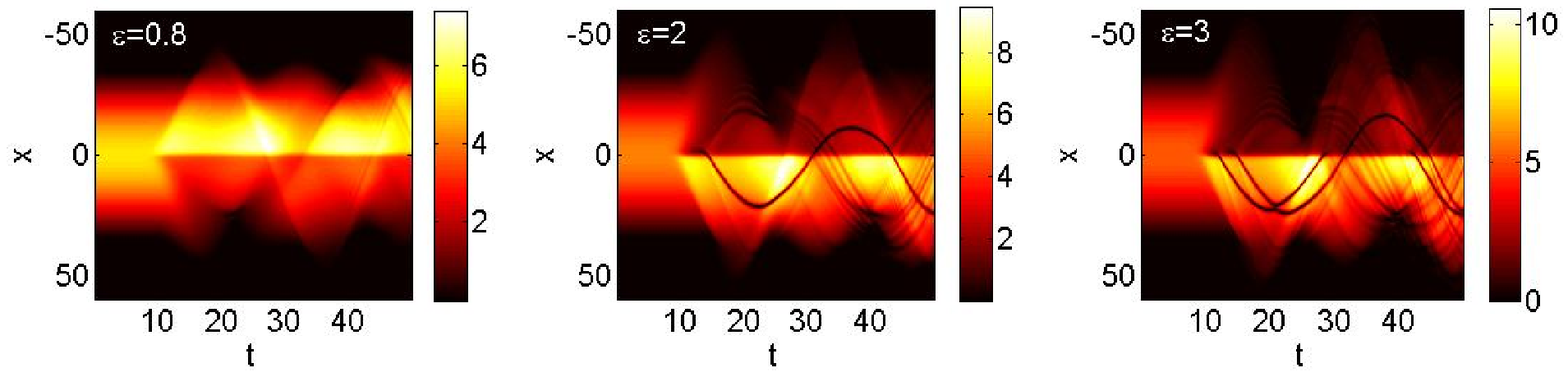}\newline
\includegraphics[width=.85\textwidth,height=.17\textheight]{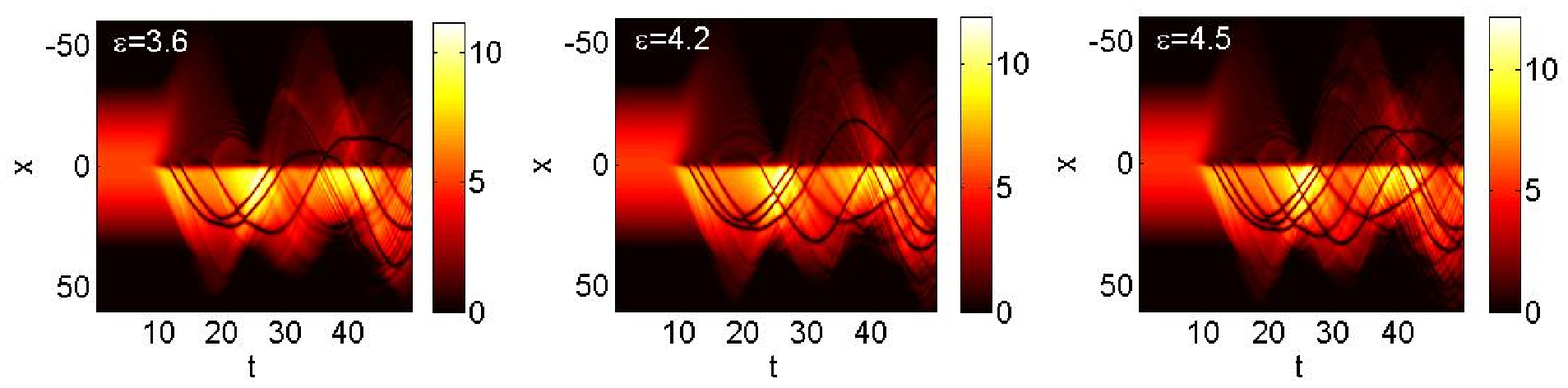}\newline
\caption{(Color online) Spatiotemporal contour plots plots, showing the evolution density $|u|^{2}$ of the condensate, as per Eq.(\ref{eq1}), with fixed chemical potential ($\mu=2.5$) and different values of the parameter $\varepsilon$. When $\varepsilon=0.8$ dark solitons are not generated, while as the value of $\varepsilon$ is increased, the formation of $1,2,3,4,5$ solitons is observed for $\varepsilon=2,3,3.6,4.2,4.5$, respectively.}
\label{fig1}
\end{figure}

In Fig.~\ref{fig1}, where some typical scenarios of the evolution are shown,
we keep the value of the chemical potential fixed, namely $\mu=2.5$, and investigate
how the value of $\varepsilon$ affects this process. We observe, at first, that dark solitons are not formed for sufficiently small values of $\varepsilon$; a pertinent example, corresponding to $\varepsilon=0.8$, is provided in the top-left panel of the figure. On the other hand, the increase of the value of $\varepsilon$ results in the formation of dark solitons. The top-middle panel of Fig.~\ref{fig1} shows that a dark soliton is emitted forwardly (i.e., toward the right portion of the BEC, with the smaller value of the scattering length) at about $t=14$ when $\varepsilon=2.0$; similarly, the top-right panel of Fig.~\ref{fig1} shows the emergence of two solitons, for $\varepsilon =3$, and so on (see a more detailed analysis below).

Figure~\ref{fig1} also demonstrates the basic features of the condensate and soliton dynamics.
In that regard, first we note that the switch-on (around $t=t_0=10$) of the external potential modifying the spatial distribution of the nonlinear coefficient, results in the excitation of the quadrupole mode of the 
condensate \cite{BECBook}. In particular, after $t=10$, the condensate shows a breathing behavior, performing the quadrupole oscillation with a characteristic frequency close to the value $\sim \sqrt{5/2} \Omega$ \cite{Delgado}. On the other hand, the dark solitons, when formed, perform oscillations due to the 
external potential. As seen in Fig.~\ref{fig1}, dark solitons with sufficiently large velocities are transmitted through the ``interface'' at the trap center ($x=0$), while lower velocity solitons are chiefly reflected by the interface. Moreover, it can readily be observed that in cases where more than two solitons are formed, they may undergo elastic collisions: see, e.g., the third panel of Fig.~\ref{fig1} (for $\varepsilon=3$), where the solitons collide 
at $t \approx 22$. Similar events arise in cases where more solitons are generated. It should also be mentioned that the soliton generation (and subsequent evolution) is followed by emission of radiation of weak linear (sound) waves, which are originally generated in the left portion of the BEC (with the larger value of the scattering length), but then travel in the whole system.

At this point, it is necessary to mention that the emergence of solitons is a continuous process.
In particular, instead of emerging only for a specific value of $\varepsilon$, the density of a dark soliton becomes deeper as $\varepsilon$ increases and, meanwhile, the phase dislocation increases also,
being associated with the presence of the dark soliton. Thus, if a dark soliton is formed for a given value of $\varepsilon$, the same soliton tends to appear earlier as $\varepsilon$ is increased.
Due to the gradual character of this process, it is difficult to quantitatively identify 
the emergence of the soliton(s). Nevertheless, our rule of identification is set in the way that,
if a dark soliton is emitted forwardly ($x>0$) before $t=25$, then it is counted. This is how we determine the number of solitons as in Fig.~\ref{fig1}.

\begin{figure}[tbp]
\centering
\includegraphics[width=.4\textwidth]{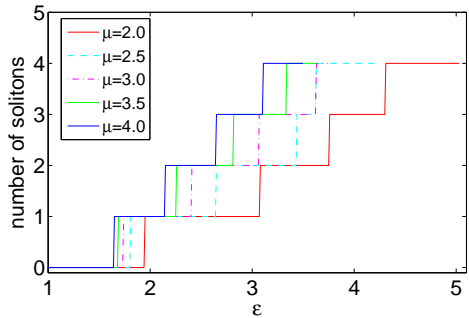}\newline
\caption{(Color online) The number of solitons emitted before $t=25$,
according to the identification rule defined in section \ref{1D},
as a step function of $\varepsilon$ for different values of the chemical potential $\mu$.
The five functions with various colors and line types represent
five different cases chosen as example,
i.e., $\mu=2.0,\,2.5,\,3.0,\,3.5\;\mathrm{and}\;4.0$.
}
\label{fig2}
\end{figure}

As mentioned before, besides $\varepsilon$, the chemical potential $\mu$ is another factor which
is important for the soliton formation. Figure~\ref{fig2} demonstrates the joint influence of these two parameters. The increase of each parameter has a positive effect on the system's capability of emitting solitons. In particular, the larger the value of $\mu$ is, the smaller the value of $\varepsilon$ at which dark solitons may be generated. For instance, applying the identification rule stated before, the second soliton emerges when $\varepsilon=3.08$ for $\mu=2.0$, or when $\varepsilon=2.41$ for $\mu=3.0$, and so on (see Fig.~\ref{fig2}).

\section{The 2D Setting}
\label{2D}

We have also studied the 2D setting, corresponding to the case of the so-called disk-shaped BECs.
%The pertinent GP equation follows the form of Eq.~(\ref{eq1}), 
%but with the spatial derivative $\partial^{2}_{x}$ being %replaced by the Laplacian $\Delta=\partial^{2}_{x}+\partial^{2}_{y}$, and the harmonic potential $V(x)$ modified to $V(r)$, where $r=x^{2}+y^{2}$, due to the nature of 2D system.
In this case, the evolution of the BEC will be described by the pertinent GP 
equation with the cubic nonlinearity \cite{BECBook,book2}:
\begin{equation}
i \partial_{t} u = -\frac{1}{2}\Delta u + V(r)u + g|u|^{2}u - \mu u,
\label{gpe2D}
\end{equation}
where $u(x,y,t)$ represents the wave function, $V(r)=(1/2)\Omega^{2}r^{2}$ is
the harmonic potential with $r^{2}=x^{2}+y^{2}$,
and the Laplacian $\Delta=\partial^{2}_{x}+\partial^{2}_{y}$ \footnote{Notice
that in this case we do not use the dynamical equation of 
\cite{gerbier,Delgado} due to the nontrivial modifications it incurs
in the presence of vortices, which constitutes a principal feature
that arises in the observed phenomenology}.
The nonlinear coefficient $g$ is assumed to take the same form, as in the 1D setting, indicating that the spatial inhomogeneity is employed along the $x$-axis.

\begin{figure}[tbph]
\centering
\includegraphics[width=.22\textwidth]{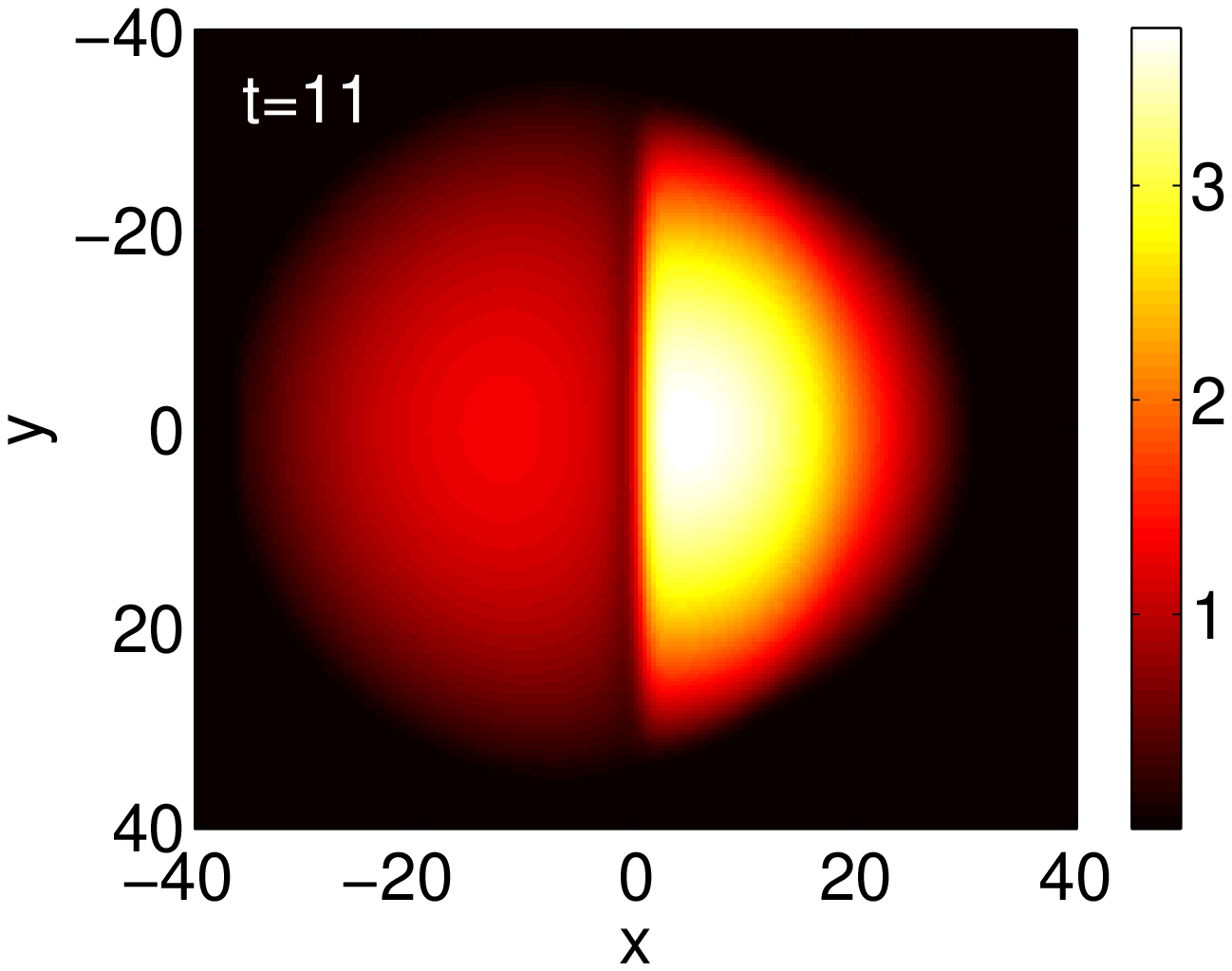}
\includegraphics[width=.22\textwidth]{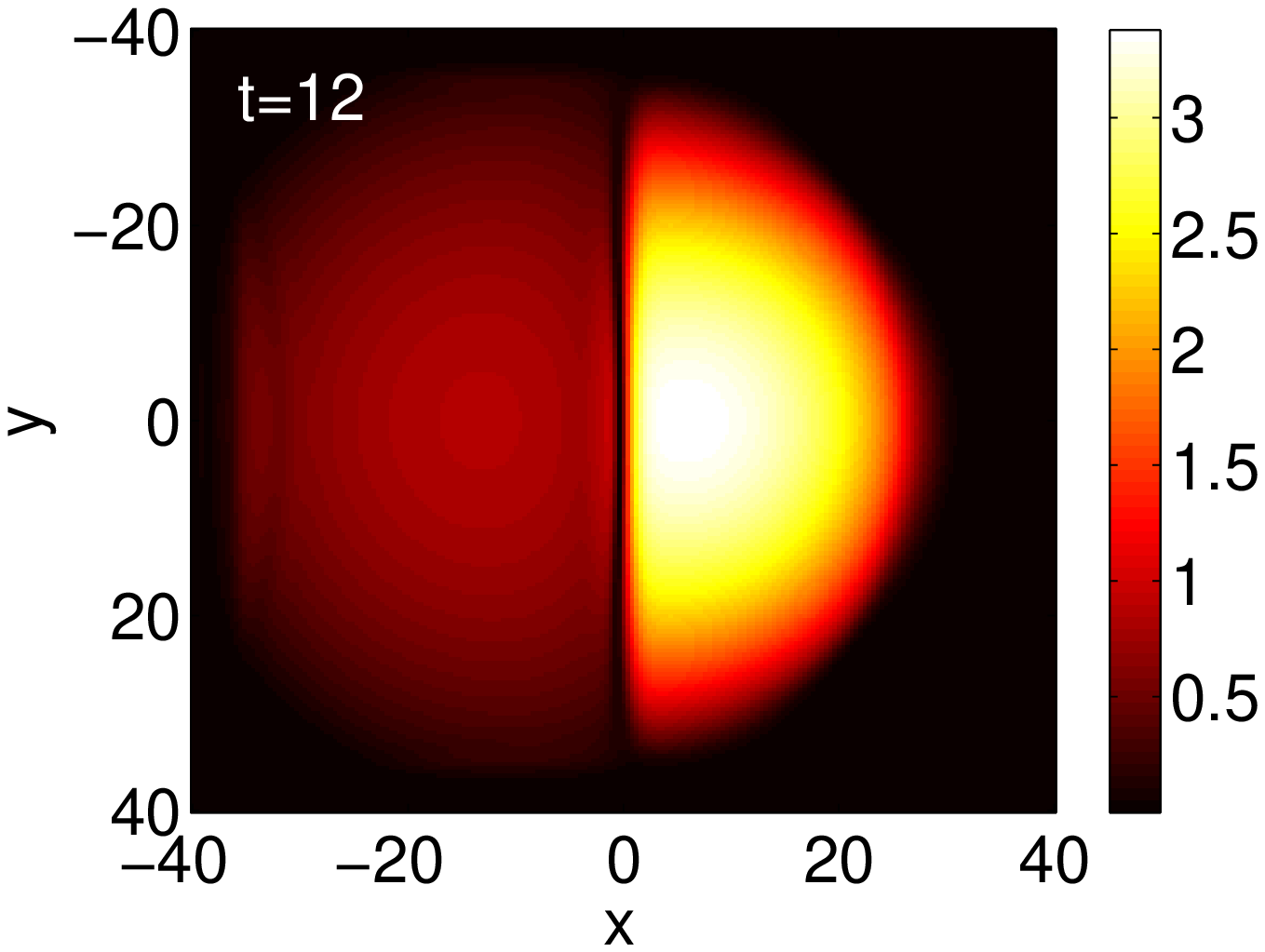}
\includegraphics[width=.22\textwidth]{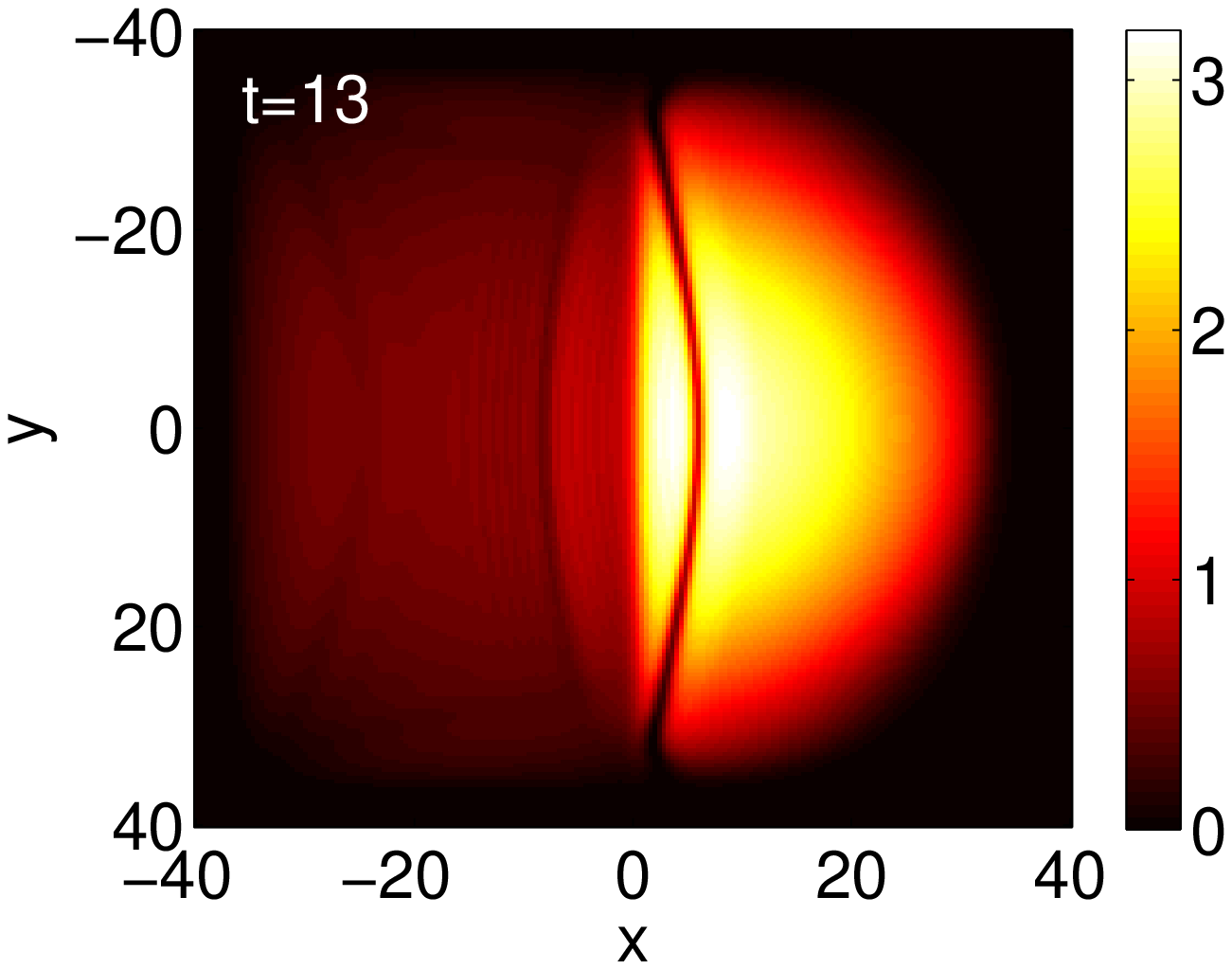}
\includegraphics[width=.22\textwidth]{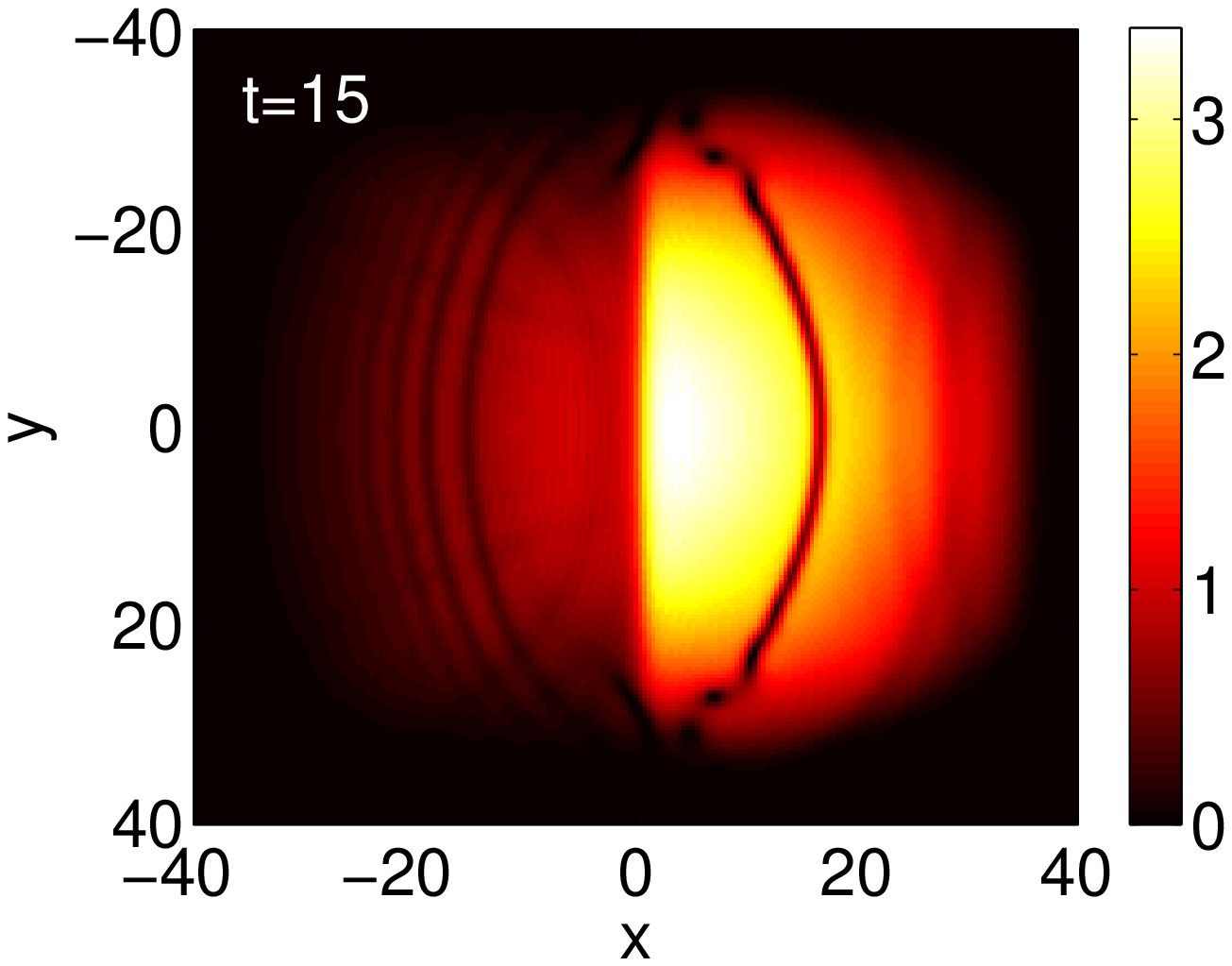}\newline
\includegraphics[width=.22\textwidth]{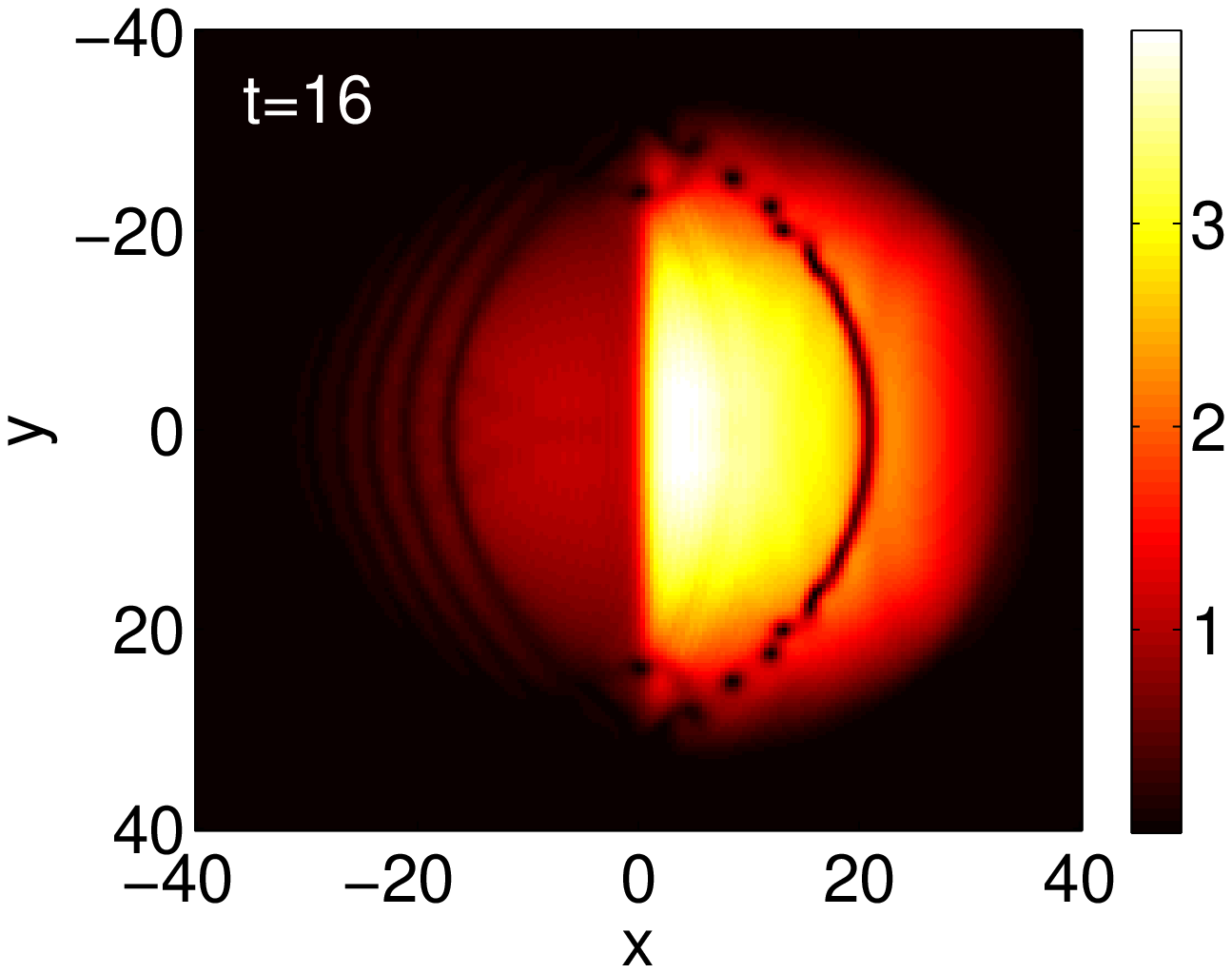}
\includegraphics[width=.22\textwidth]{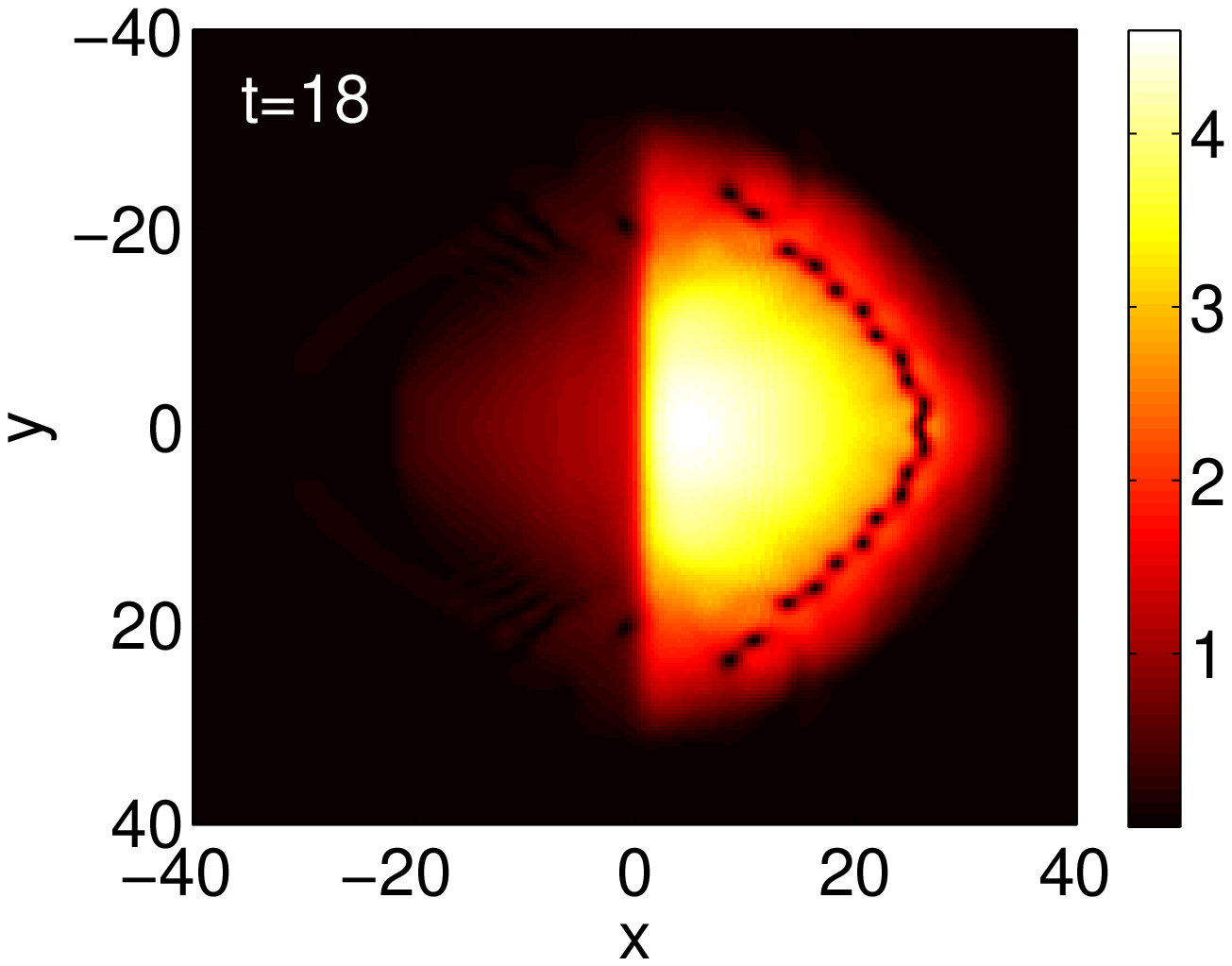}
\includegraphics[width=.22\textwidth]{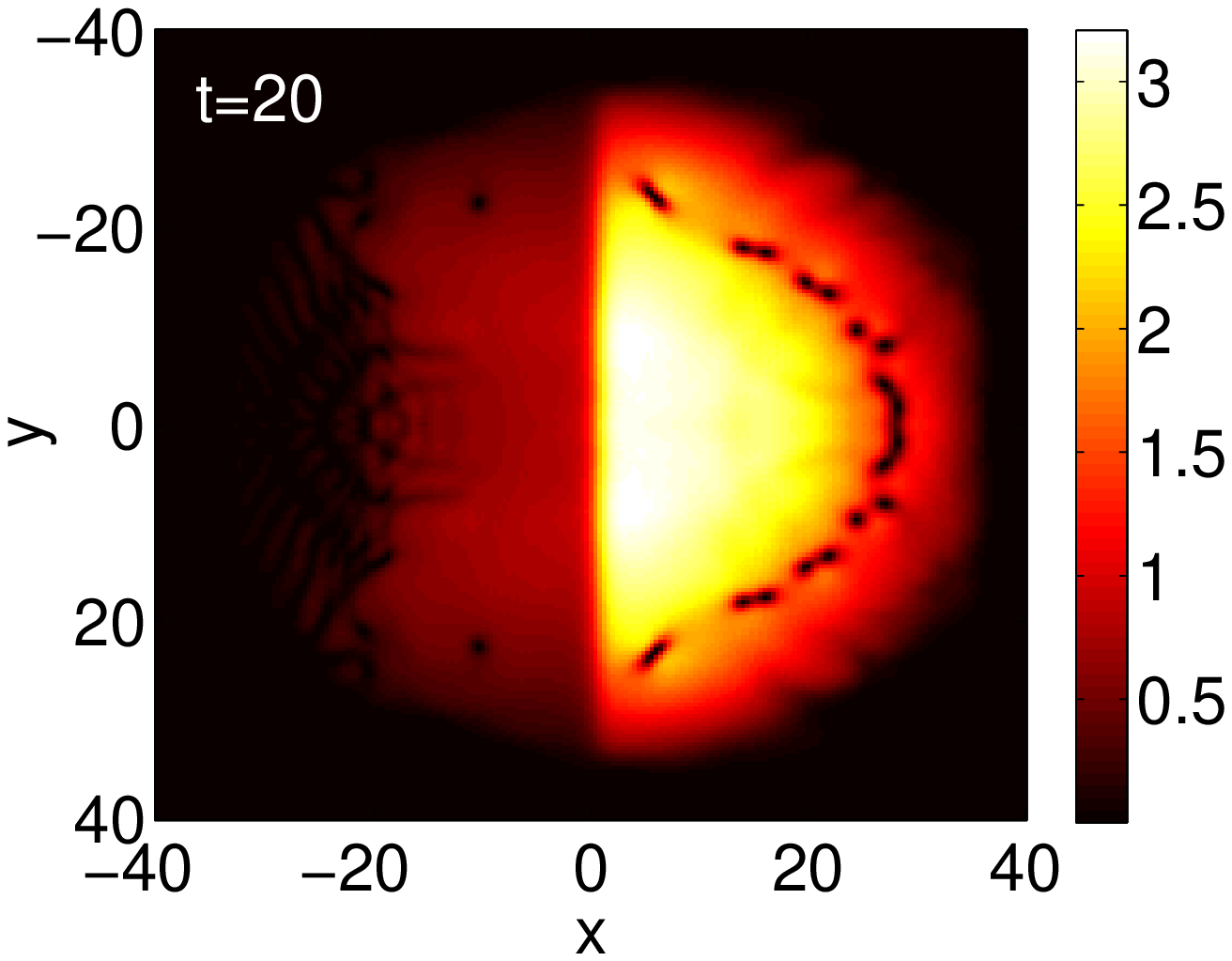}
\includegraphics[width=.22\textwidth]{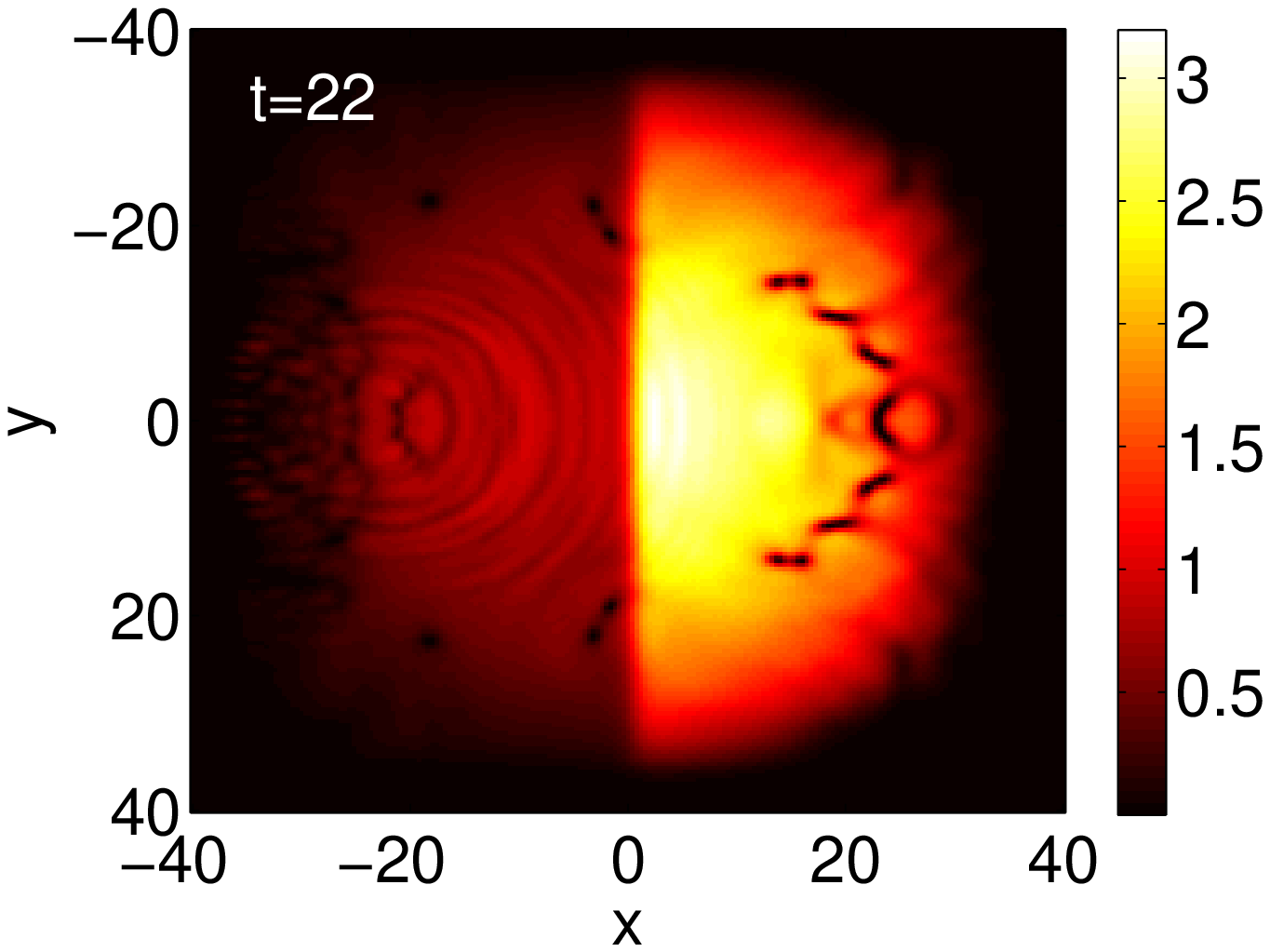}\newline
\includegraphics[width=.22\textwidth]{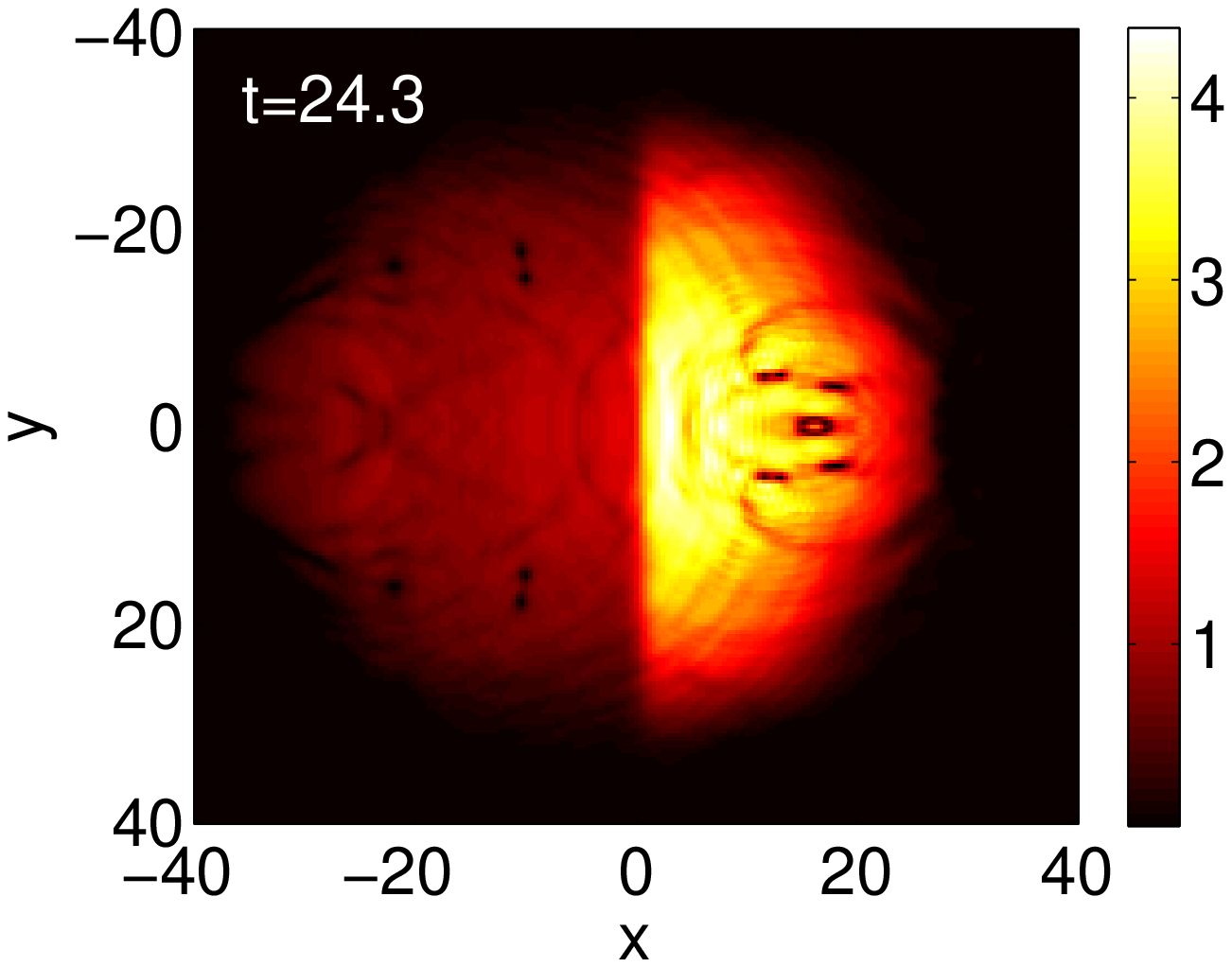}
\includegraphics[width=.22\textwidth]{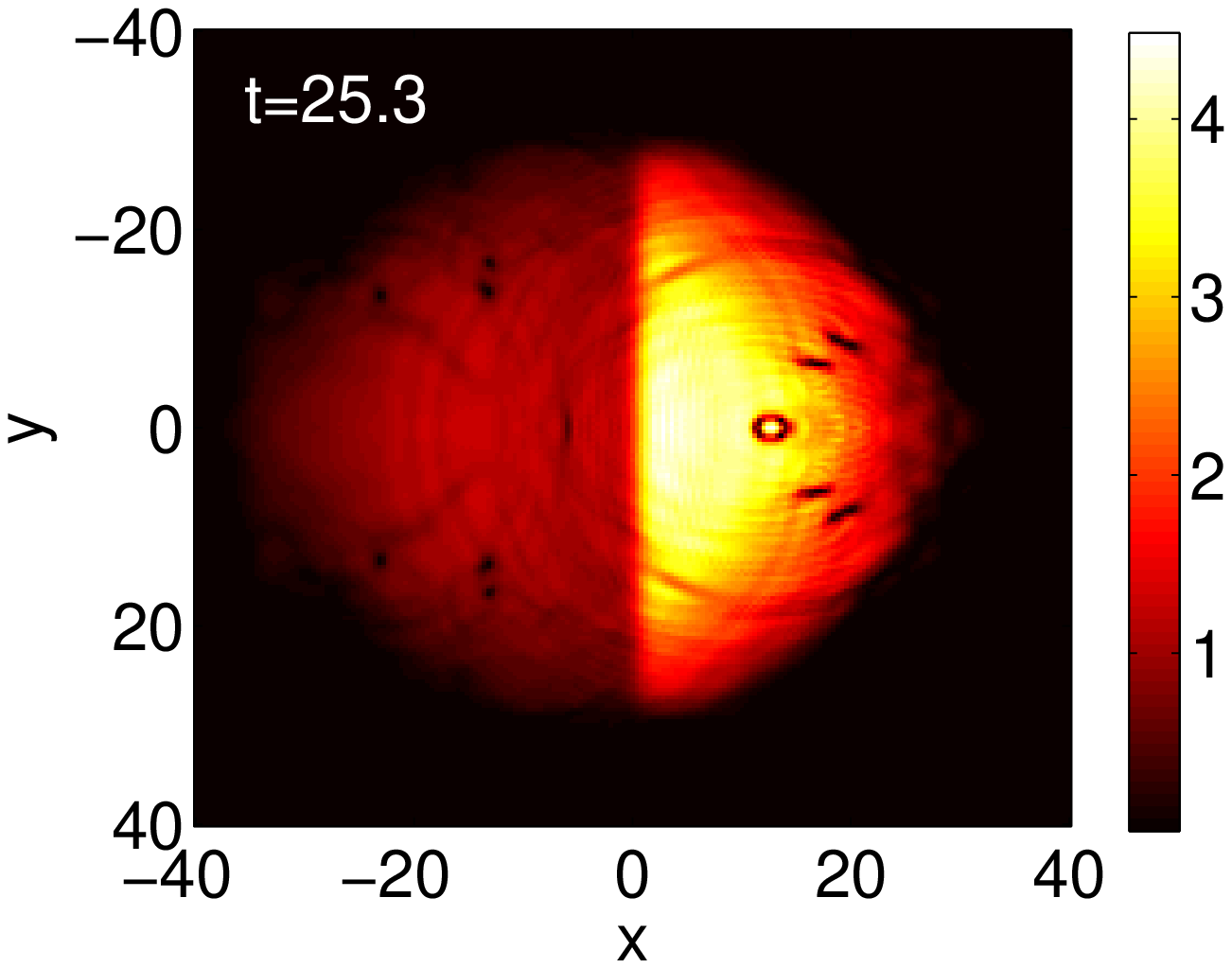}
\includegraphics[width=.22\textwidth]{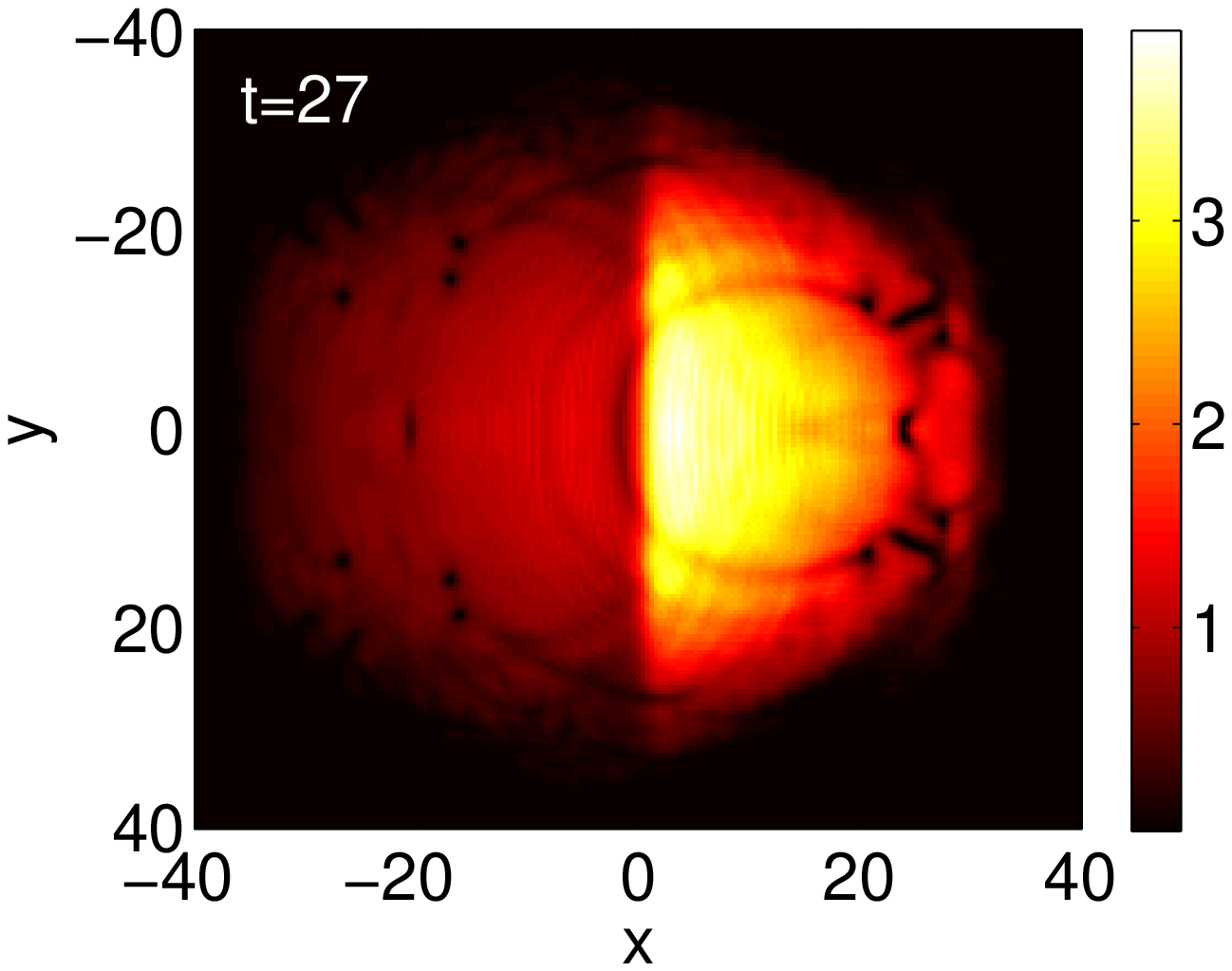}
\includegraphics[width=.22\textwidth]{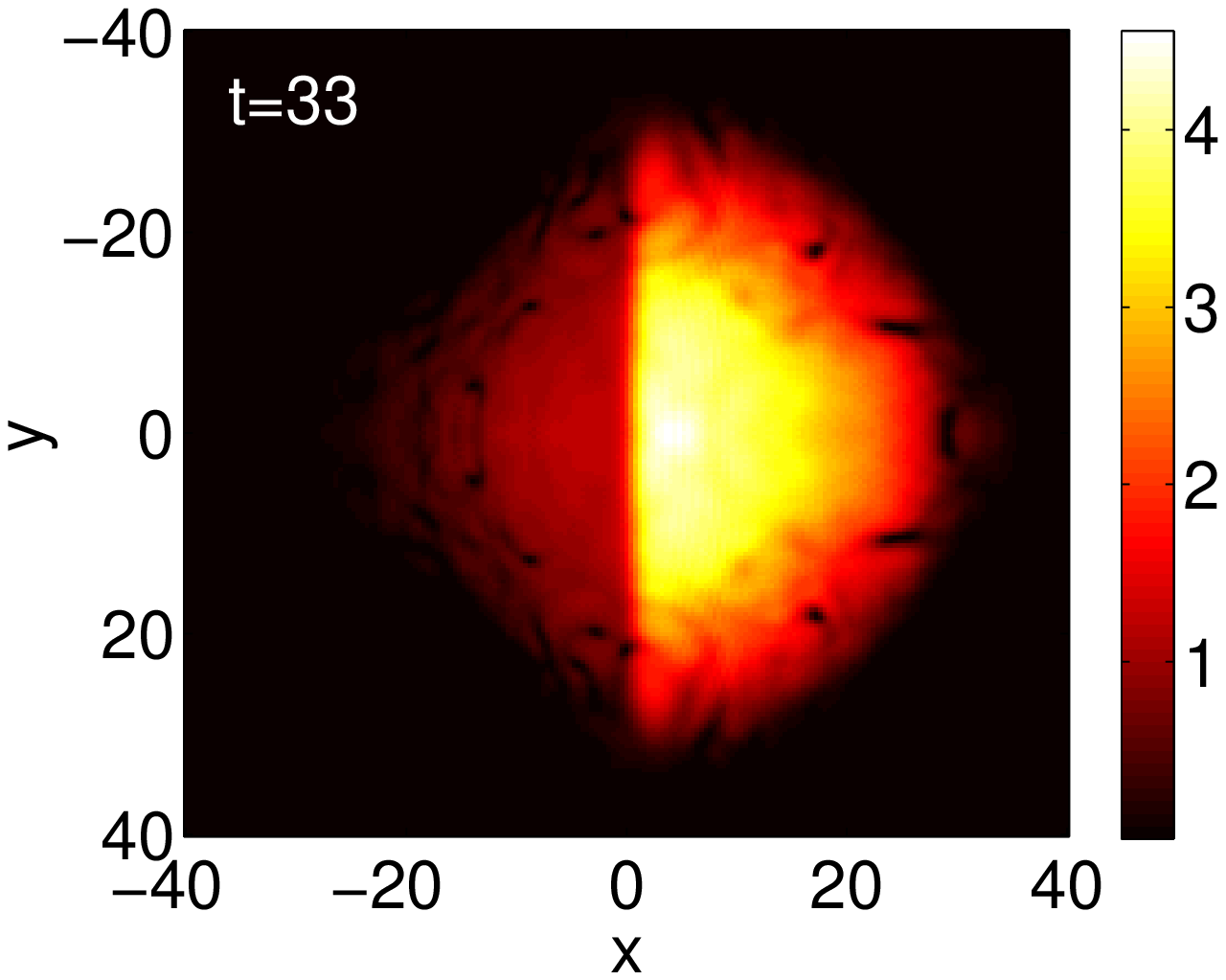}\newline
\caption{(Color online) Contour plots showing snapshots of the evolution of the
2D condensate density for $\Omega=0.08$ and $\varepsilon=2.5$ (the chemical potential is $\mu=3$).}
\label{fig3}
\end{figure}

As before, we start the simulations with the steady state of the system, corresponding to $g=1$.
The parameters $s$, $t_{0}$ and $\tau$ remain the same as in the 1D setting.
During the evolution, dark solitons are emitted when the amplitude $\varepsilon$ and
chemical potential $\mu$ are sufficiently large. Moreover, another parameter is found to play an important role in this process, namely the normalized strength $\Omega$ of the trapping potential $V(r)$.
In the numerical results of this section, $\mu$ is fixed to $3.0$, as its influence
on the dark soliton has been discussed in detail in the 1D case, and we focus on cases
with various pairs of $\Omega$ and $\varepsilon$. Figure~\ref{fig3}
describes a typical evolution scenario, by showing snapshots of the BEC density
for $\Omega=0.08$ and $\varepsilon=2.5$. Starting with a stationary
ground state (similarly to the 1D case), the condensate becomes anisotropic due to the change of the value of the nonlinear coefficient;
a rectilinear dark soliton is generated near the line $x=0$,
and then it moves to the forward direction ($x>0$), as shown in the top panels of the figure.
Subsequently, this rectilinear dark soliton does not oscillate as it does in the 1D setting;
instead, it undergoes a bending and, after moving forward for a short time, it finally breaks up
into a sequence of vortex pairs, as a result of the onset of the snaking instability (see, e.g., \cite{djf} and references therein, as well as \cite{bpa} for relevant experimental observations) -- see
the middle row of Fig.~\ref{fig3}. The generated vortices are moving due to 
their interactions and the action of the external potential, and some of them disappear at later times. It is also worth observing (see the bottom row of Fig.~\ref{fig3}) the manifestation of an interesting transient phenomenon, namely the formation of a closed-loop structures, reminiscent of ring dark solitons in BECs \cite{rds} which, however, eventually disappear. After a certain time, no specific pattern is captured in the density plot, as shown in the last panel of the figure. 

\begin{figure}[tbph]
\centering
\includegraphics[width=.3\textwidth]{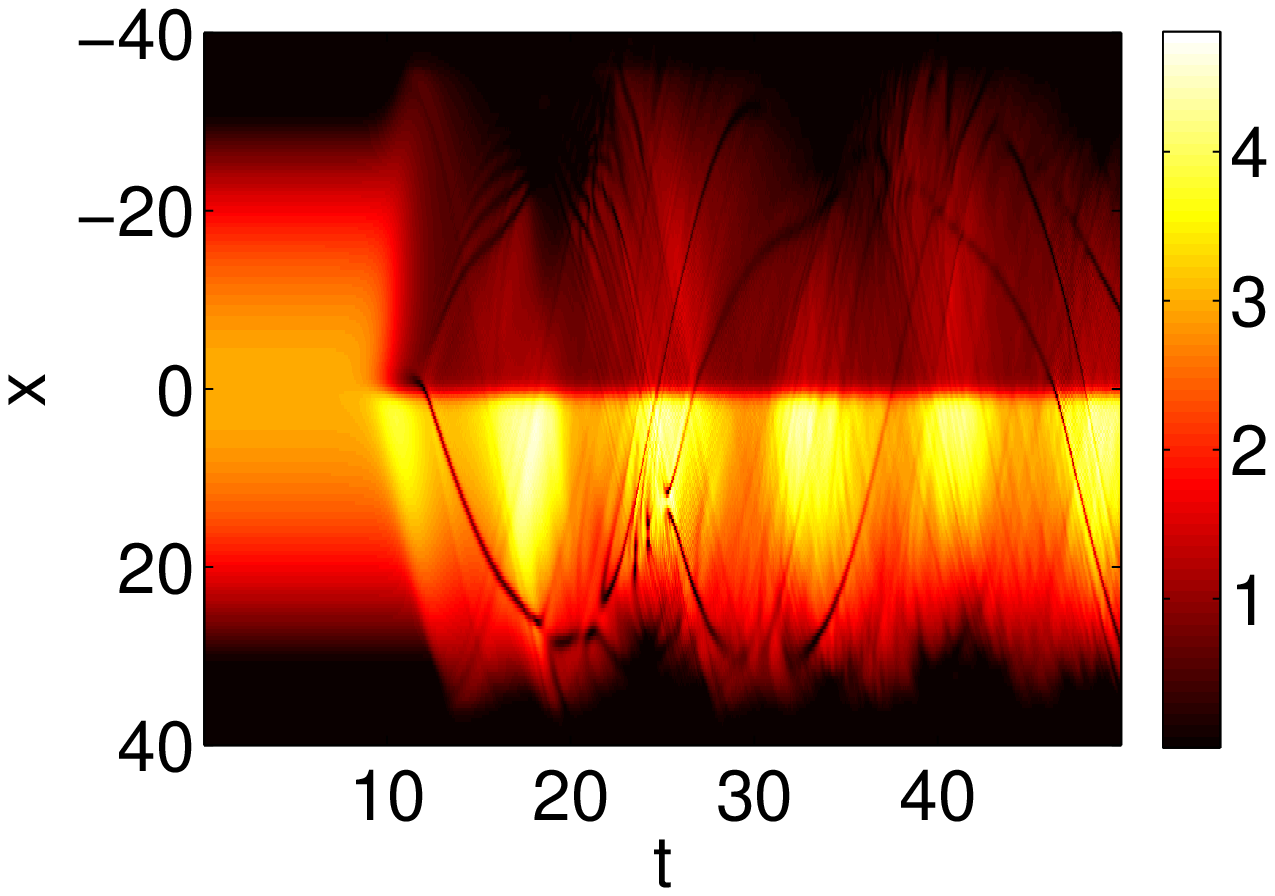}
\includegraphics[width=.26\textwidth]{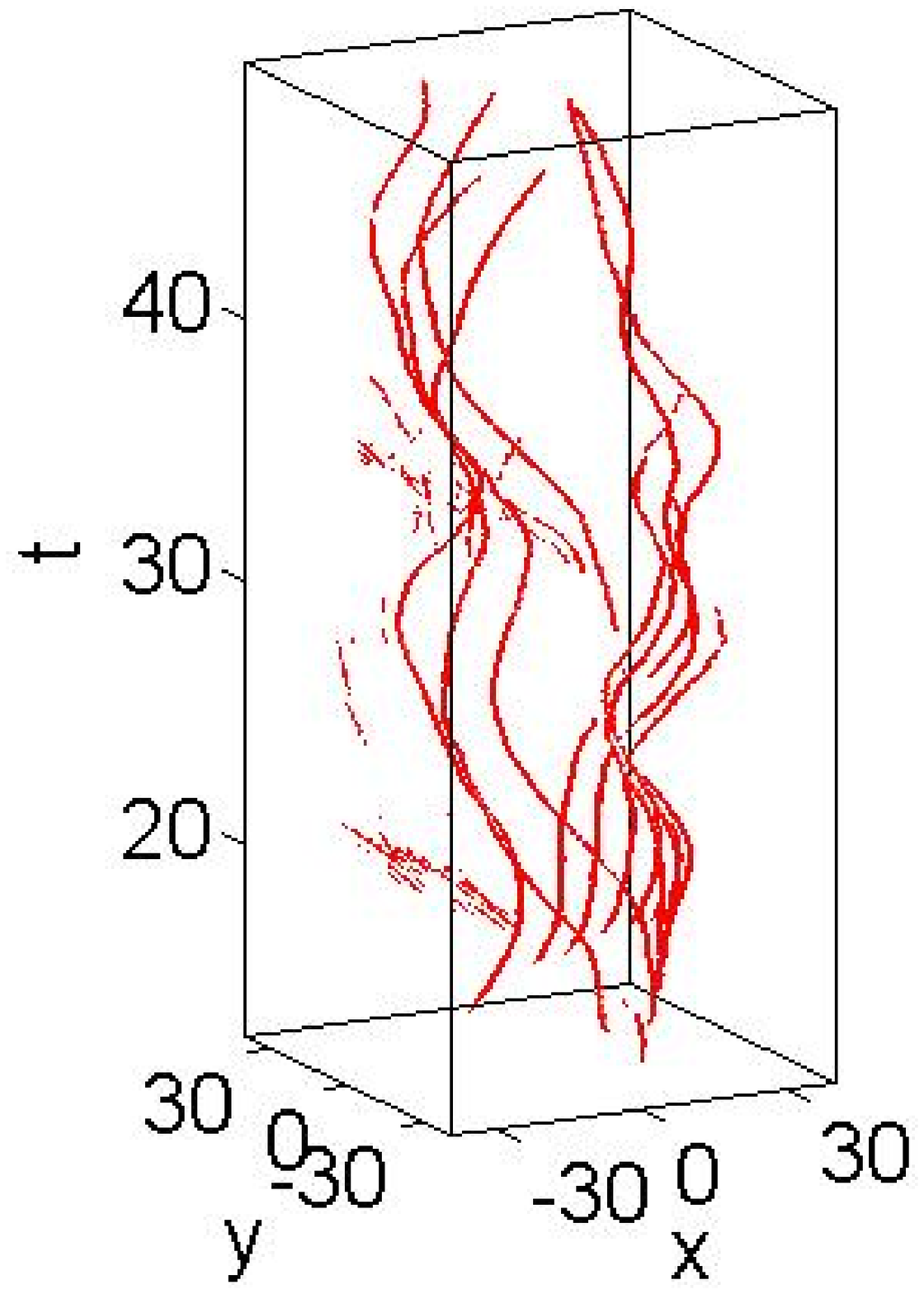}\newline
\includegraphics[width=.3\textwidth]{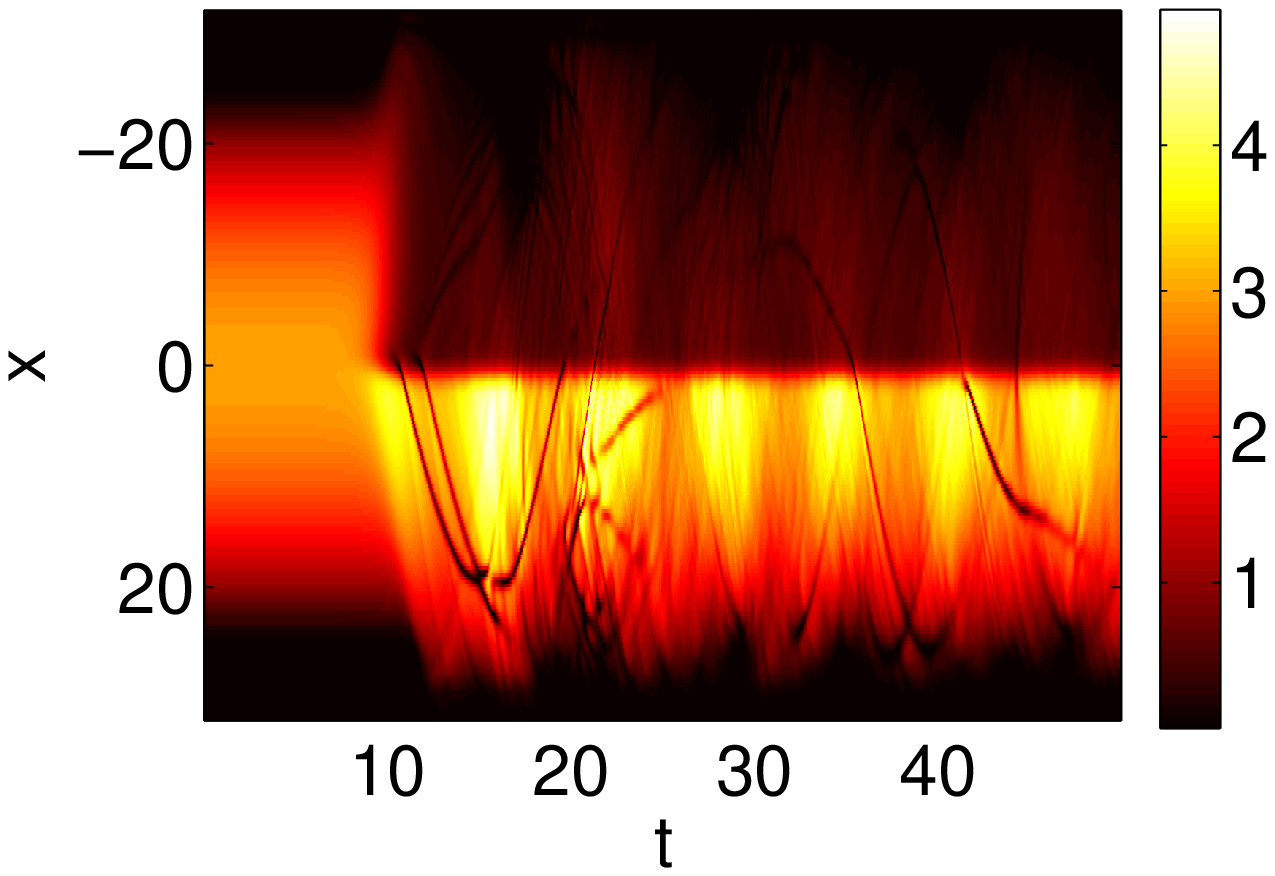}
\includegraphics[width=.26\textwidth]{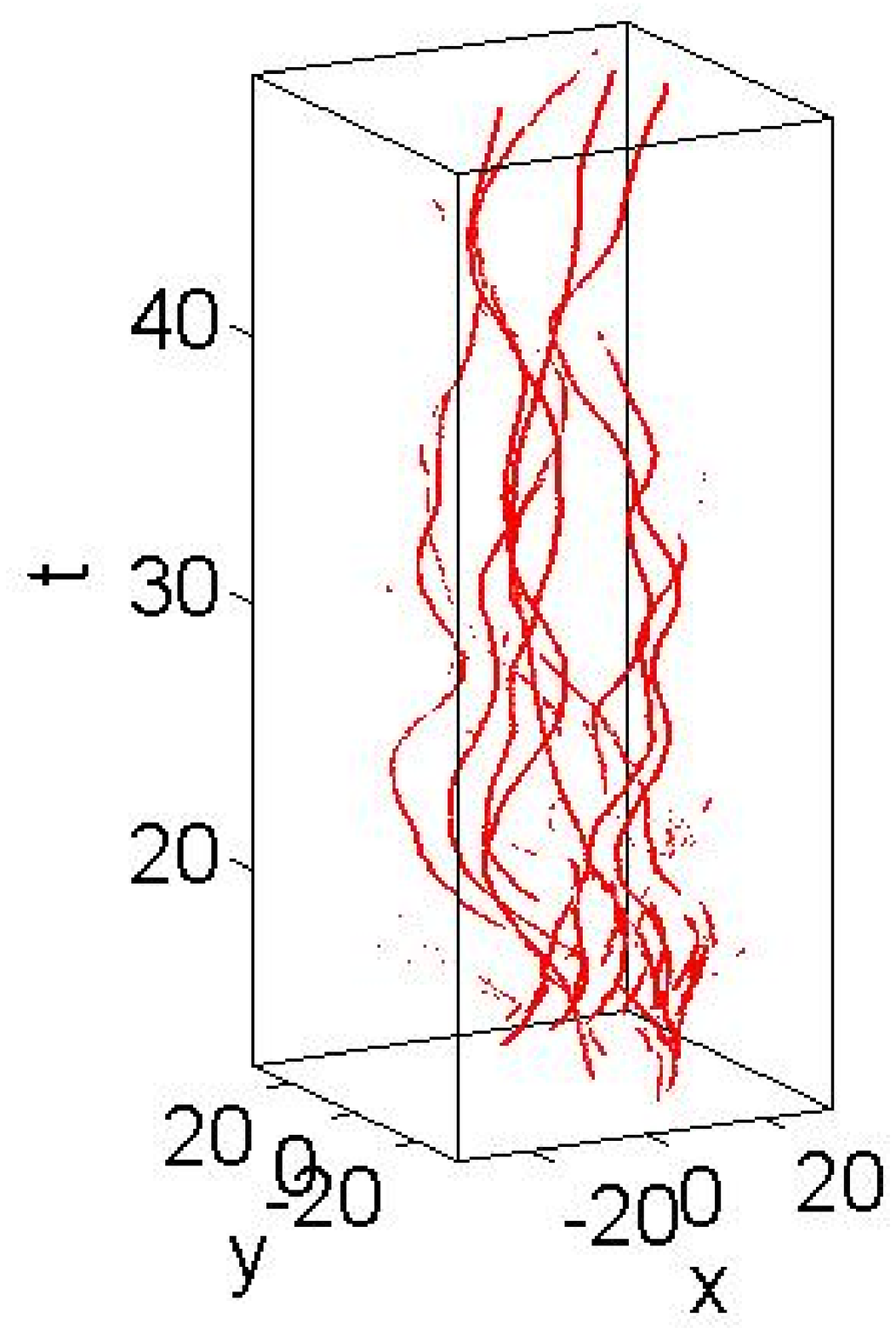}\newline
\includegraphics[width=.3\textwidth]{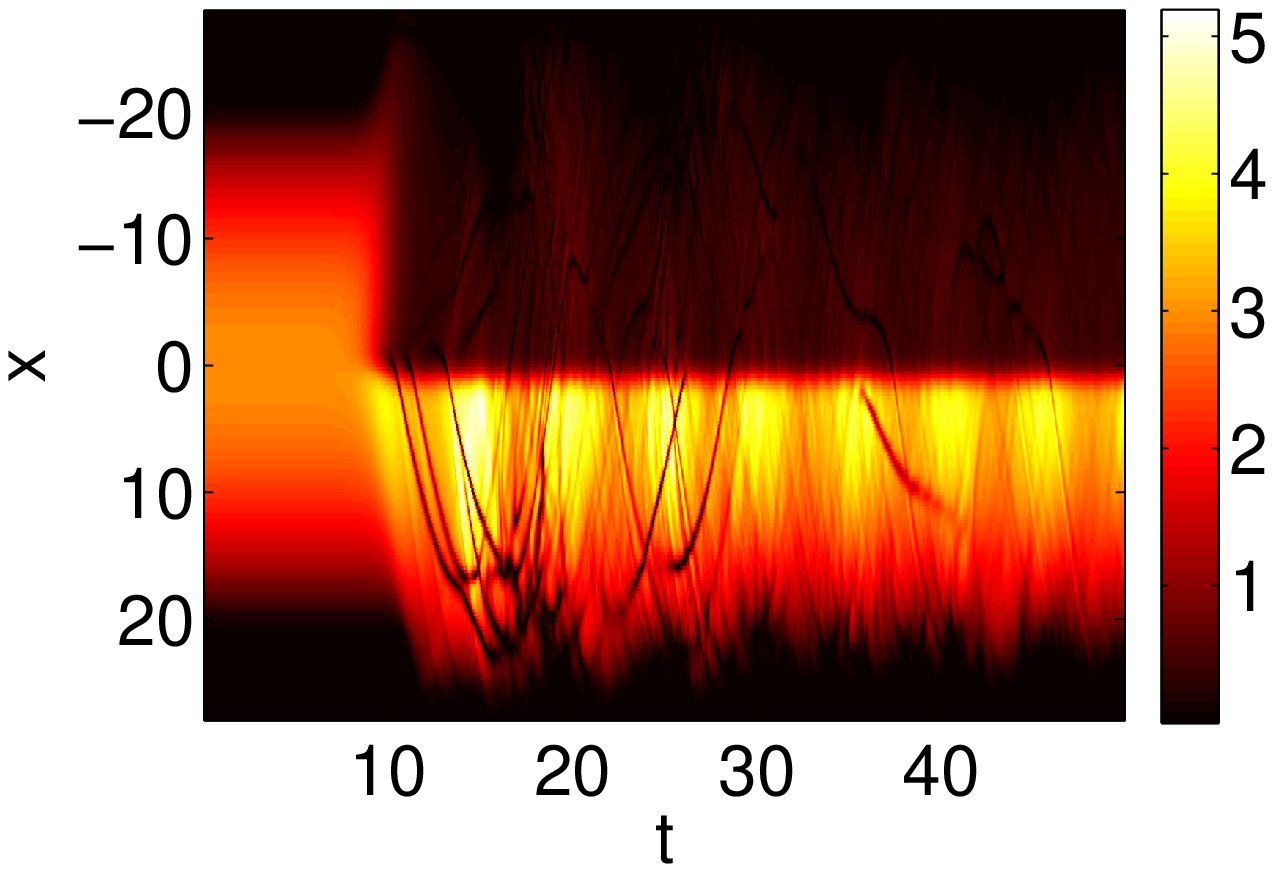}
\includegraphics[width=.26\textwidth]{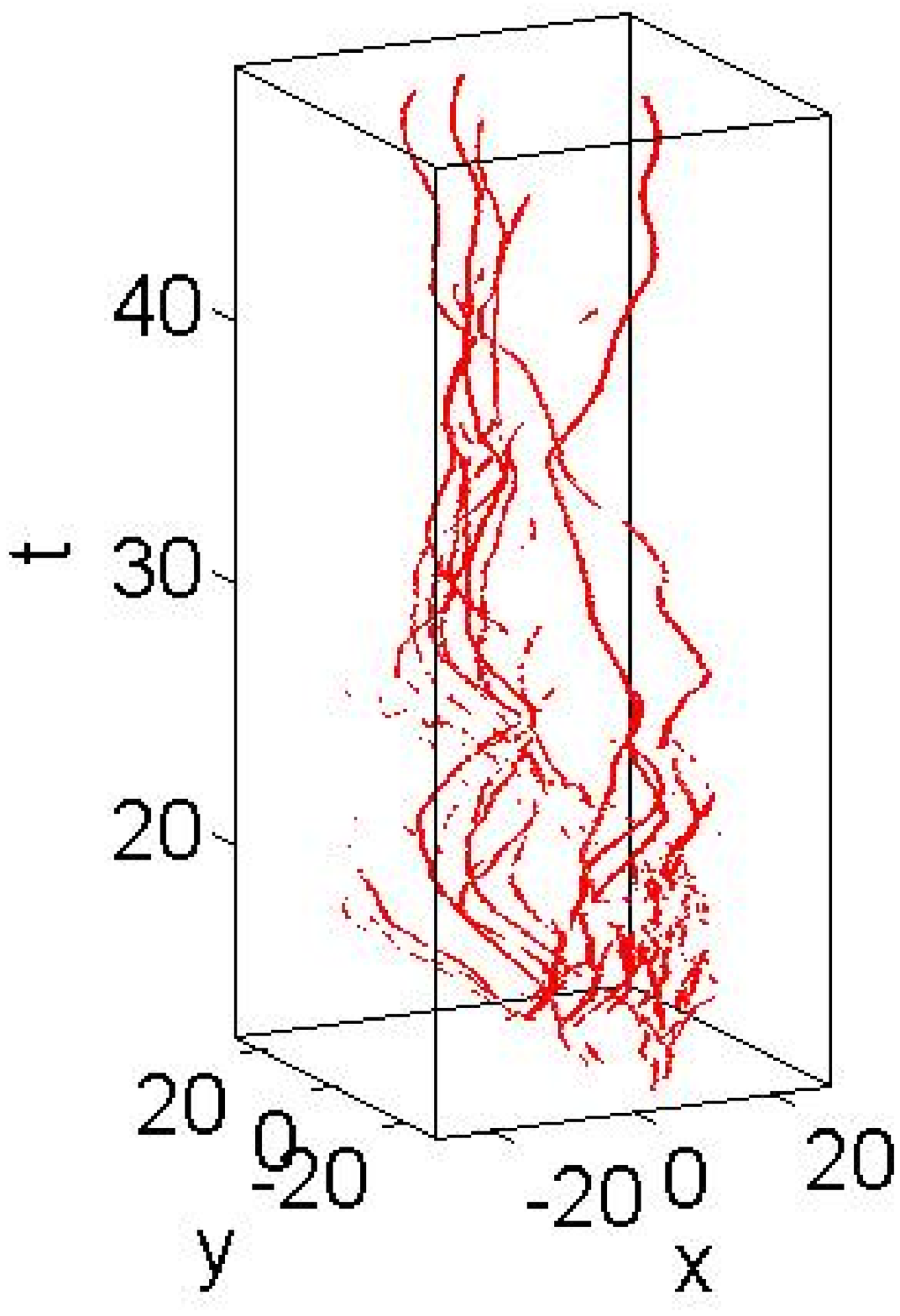}\newline
\caption{(Color online) Contour plots showing the evolution of the density on the
$y=0$ plane (left panels) and the spatiotemporal contour plots of the vorticity (right panels) -- see text.
From top to bottom, 
parameter values are $\Omega=0.08,\,0.1\;\mathrm{and}\;0.12$,
$\varepsilon=2.5,\,5.0\;\mathrm{and}\;7.5$, respectively (the chemical potential is $\mu=3$).
}
\label{fig4}
\end{figure}

In order to monitor the overall evolution of the vortices in the system,
we compute the vorticity based on the corresponding fluid velocity (see, e.g., Ref.~\cite{jackson}), namely, %
\begin{equation}
\mathbf{v}_{s} = -i \frac{u^{*}\mathbf{\nabla}{u} - u \mathbf{\nabla}{u^{*}}}{|u|^{2}}.
\label{eq4}
\end{equation}
The fluid vorticity is then defined as
\begin{equation}
\mathbf{\omega} = \mathbf{\nabla} \times \mathbf{v}_{s}.
\label{eq5}
\end{equation}
According to the above, the evolution of the vortex structures depicted in Fig.~\ref{fig3} is
presented in the top right panel of Fig.~\ref{fig4}. The accompanying top left panel of the same figure is a spatiotemporal contour plot showing the evolution of the condensate density along the $x$-direction (i.e., on the plane $y=0$), while the middle and bottom panels correspond to different (larger) values of the parameters $\Omega$ and $\varepsilon$. This figure provides a view 
complementary to the one of Fig.~\ref{fig3} concerning the emergence of the
dark soliton and vortices, and their subsequent evolution. It is interesting
to highlight that while, as indicated above, no specific density pattern
seems to be selected by the asymptotic dynamics, nevertheless, it is
clear that the topological charge (ultimately) induced by the nonlinearity
step clearly seems to persist throughout the timescales that monitored herein.

\begin{figure}[tbph]
\centering
\includegraphics[width=.22\textwidth]{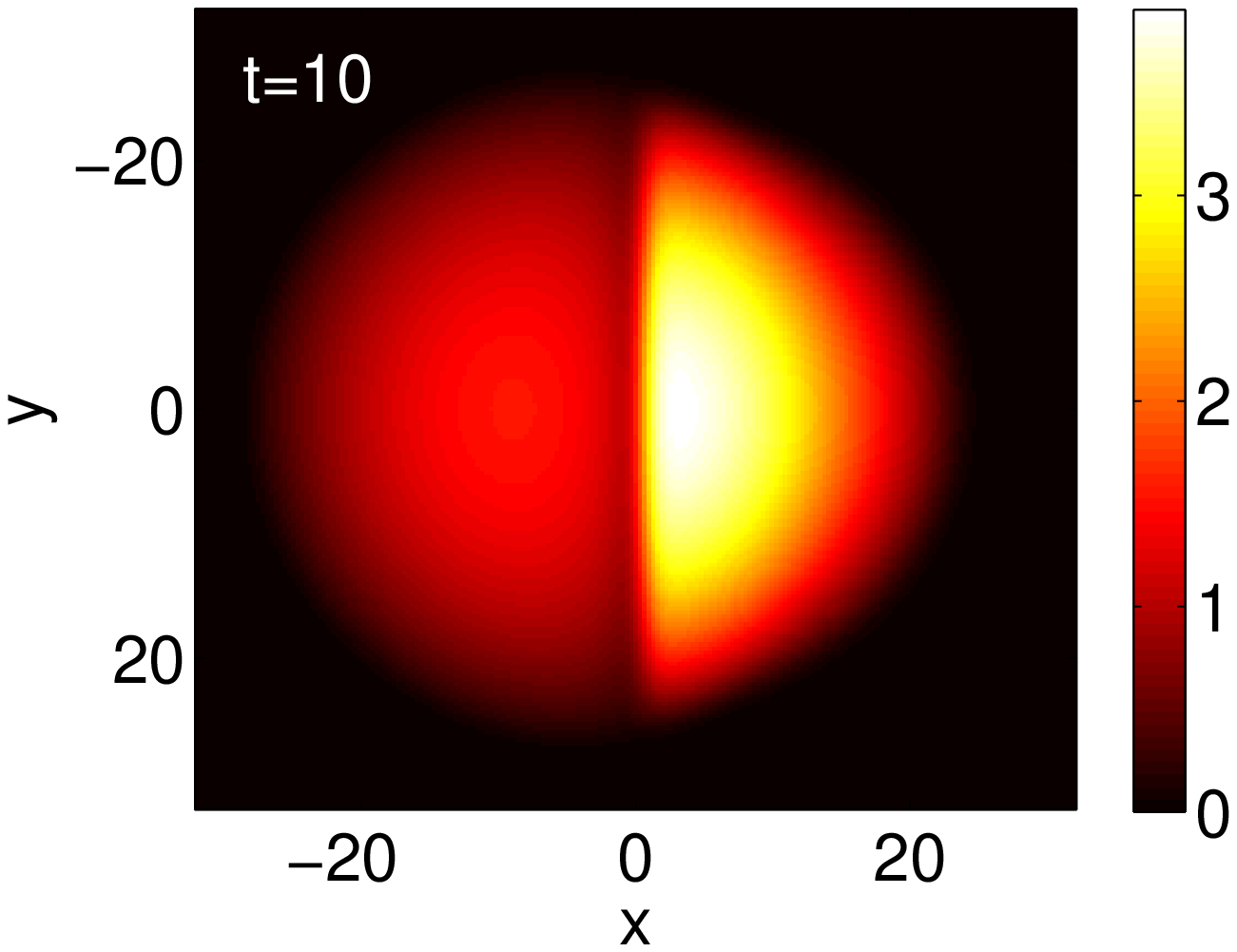}
\includegraphics[width=.22\textwidth]{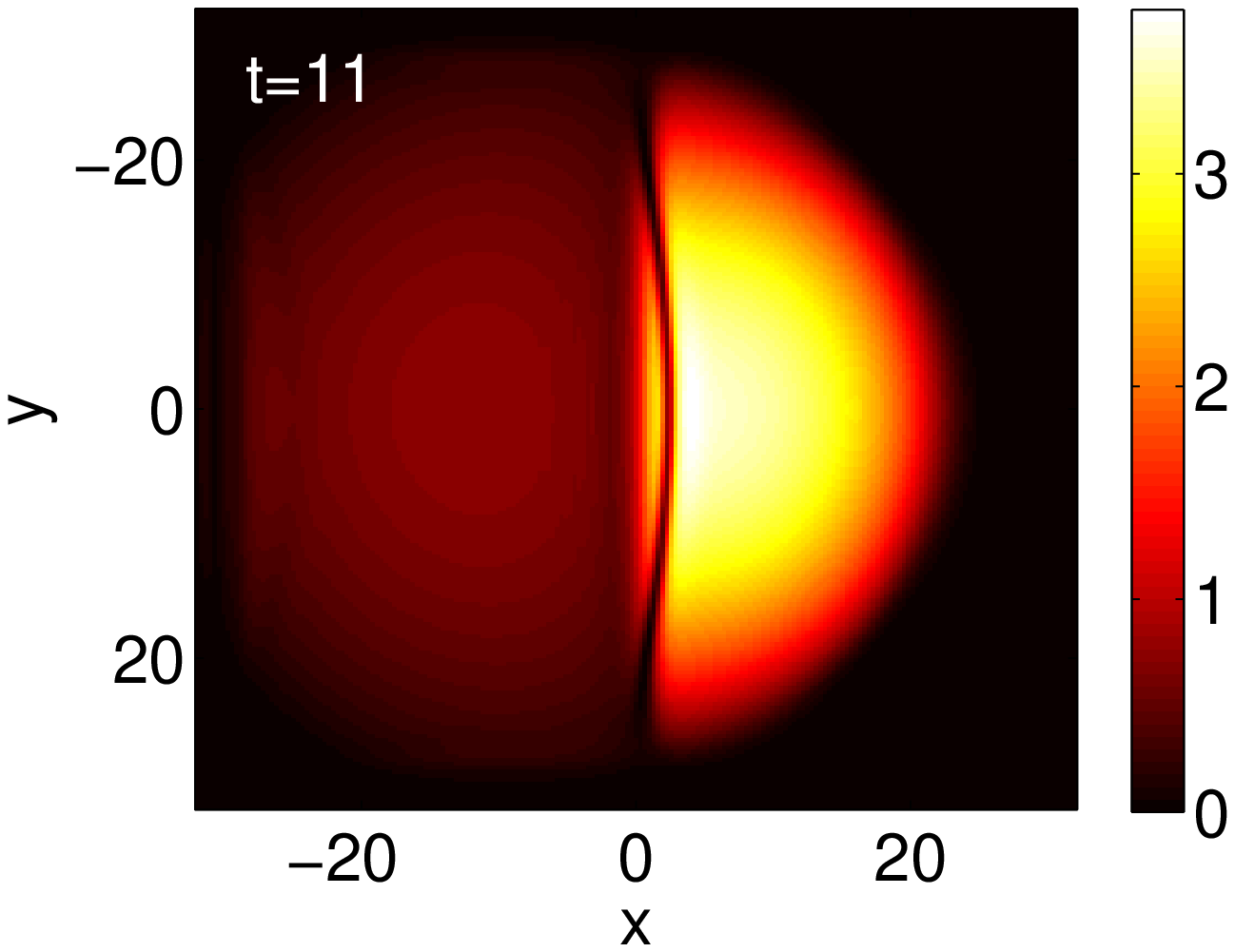}
\includegraphics[width=.22\textwidth]{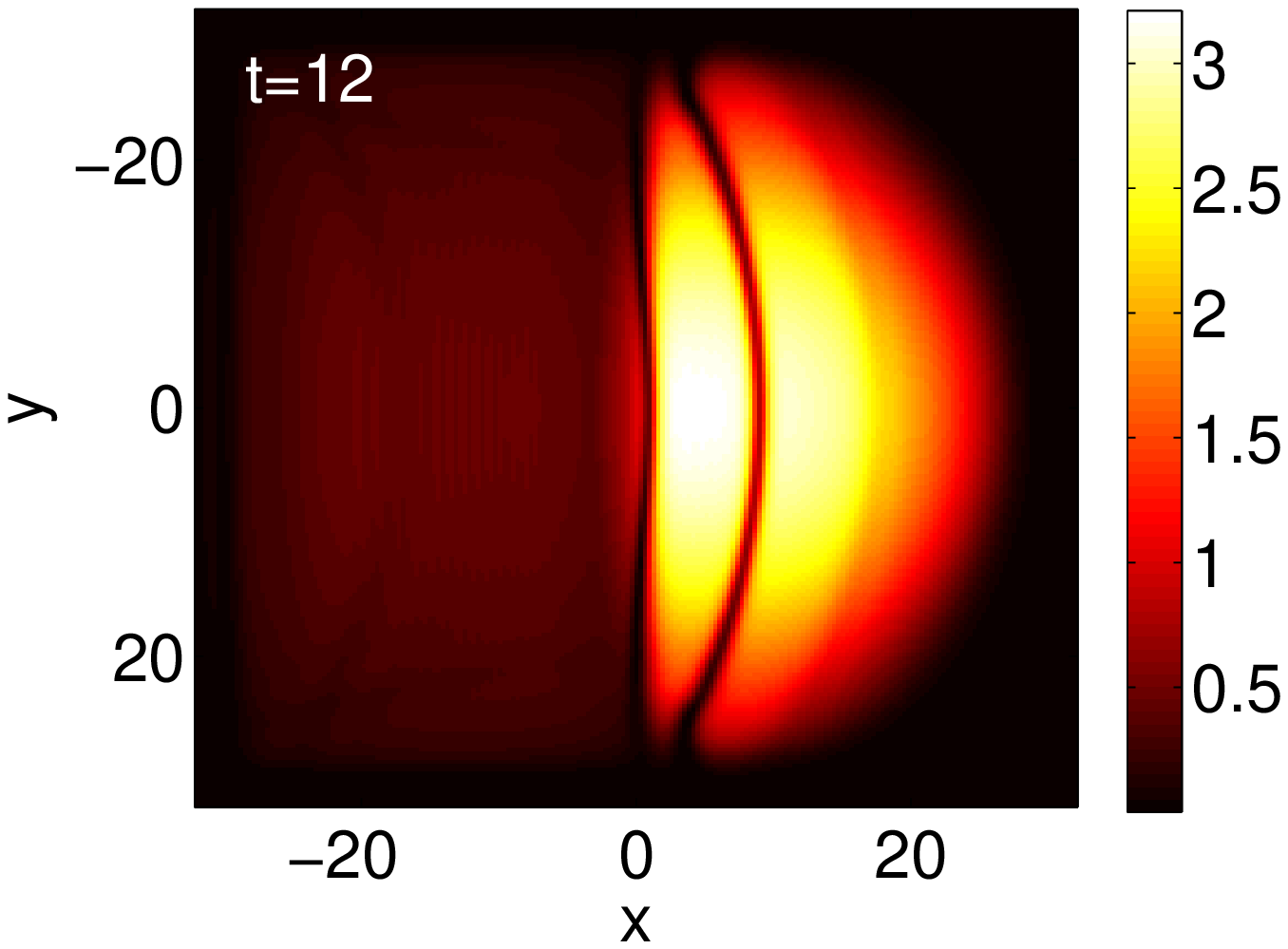}
\includegraphics[width=.22\textwidth]{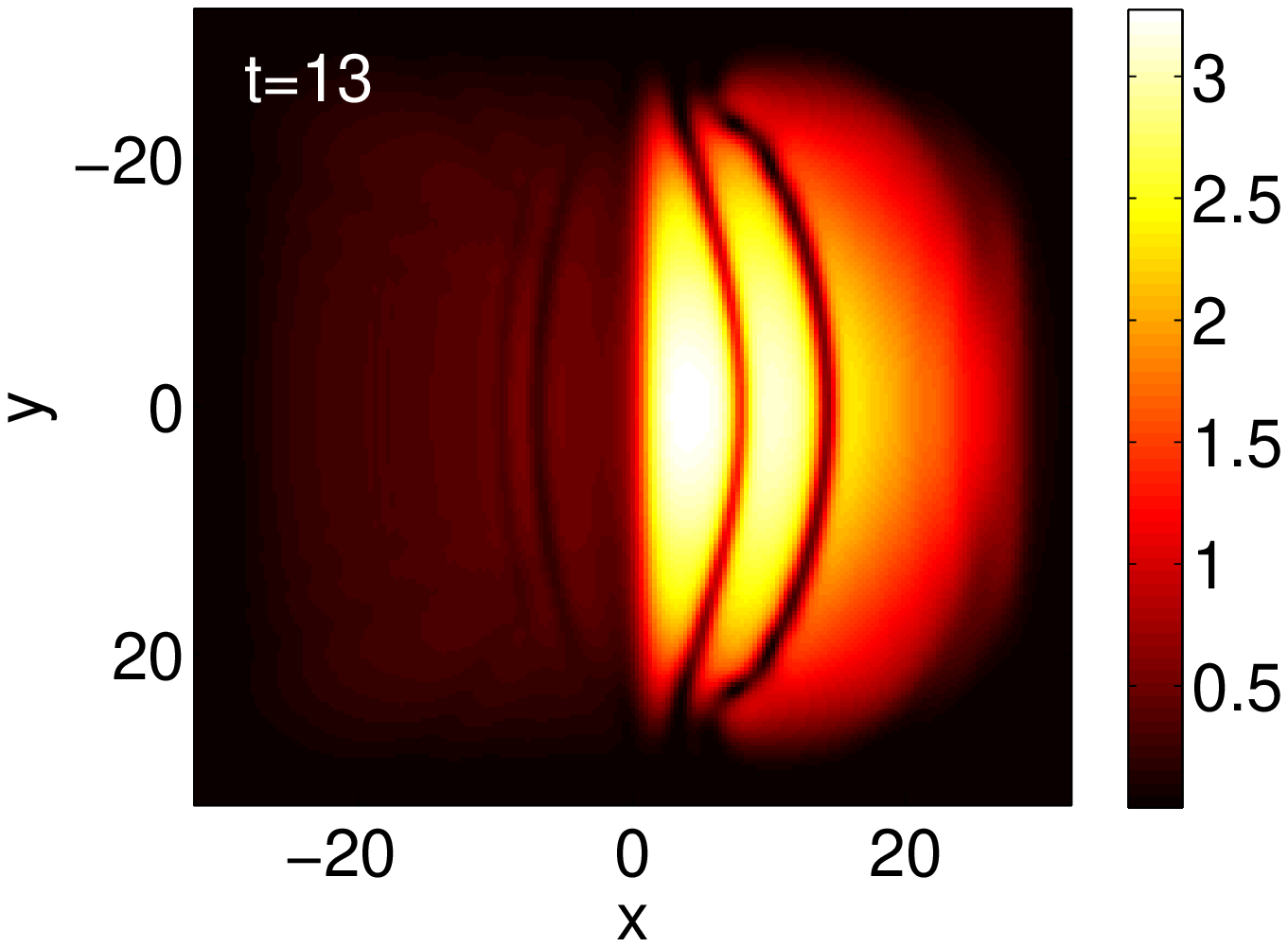}\newline
\includegraphics[width=.22\textwidth]{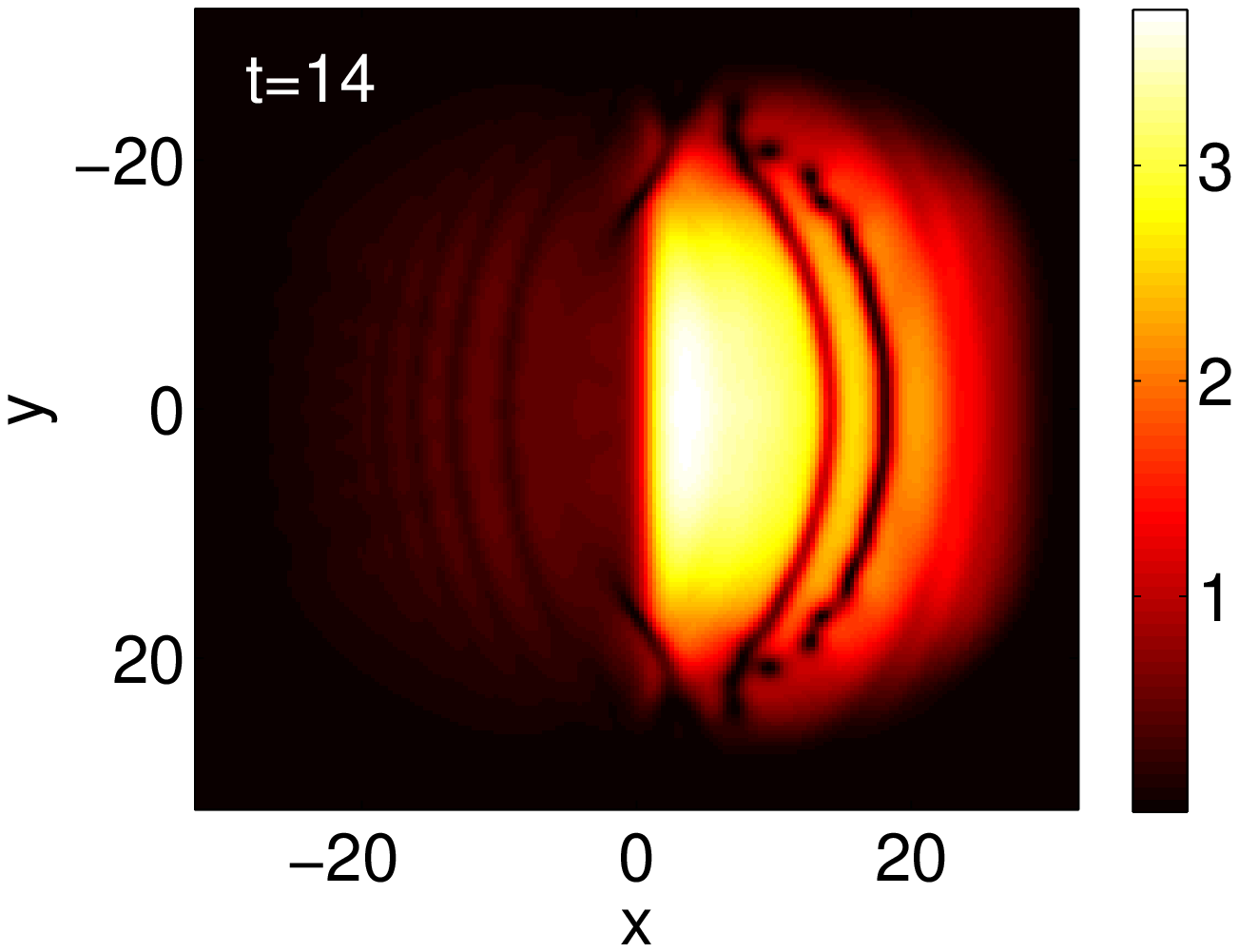}
\includegraphics[width=.22\textwidth]{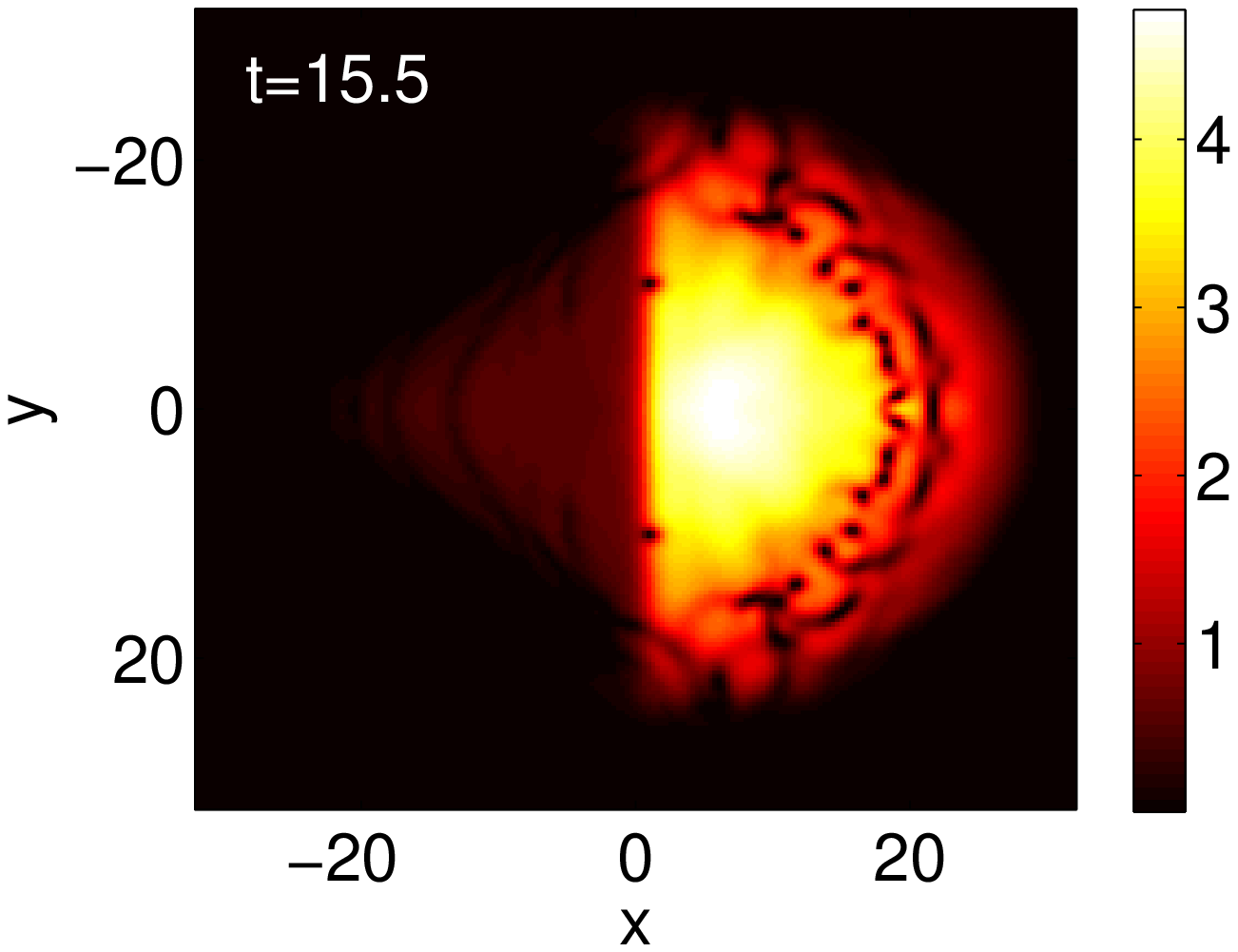}
\includegraphics[width=.22\textwidth]{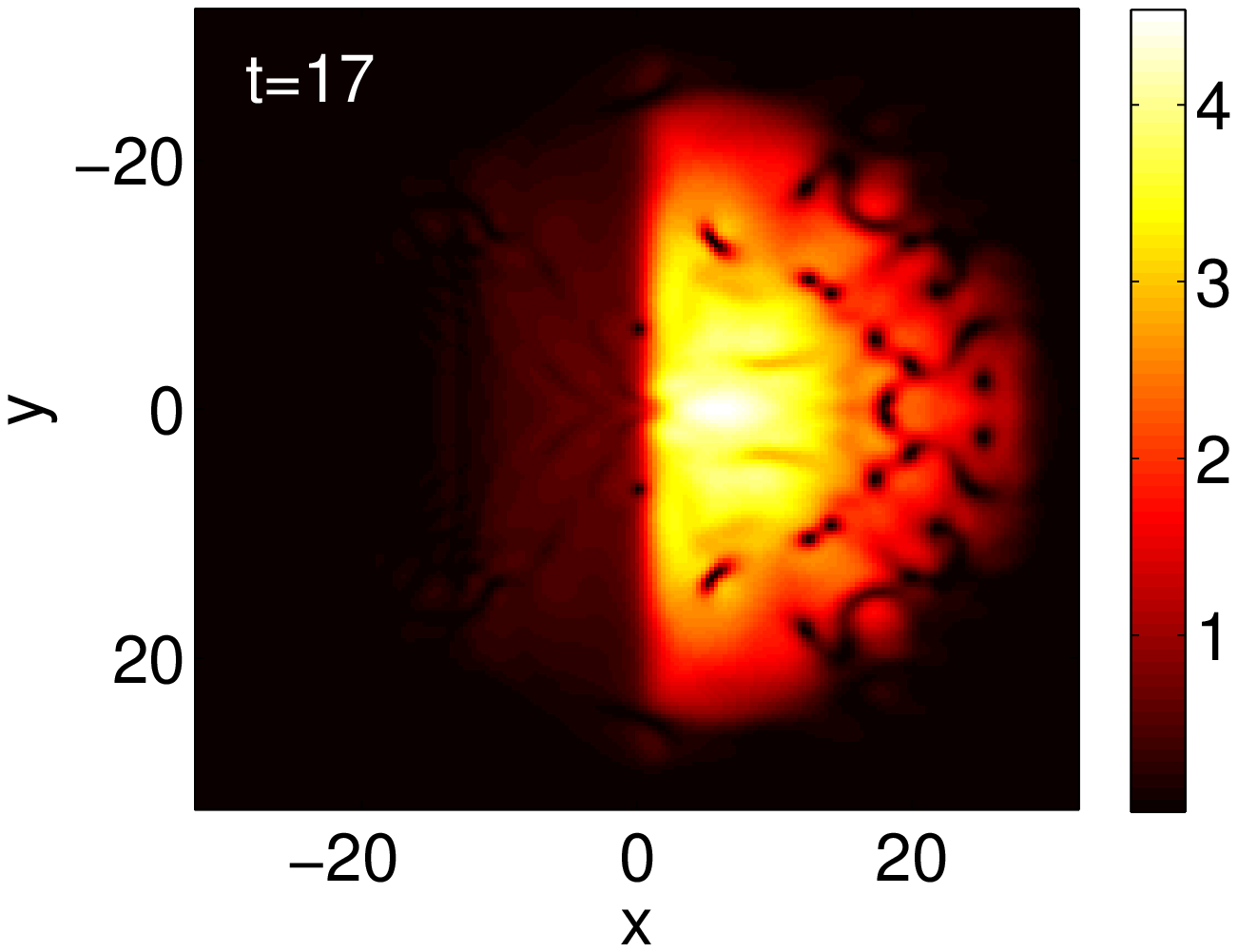}
\includegraphics[width=.22\textwidth]{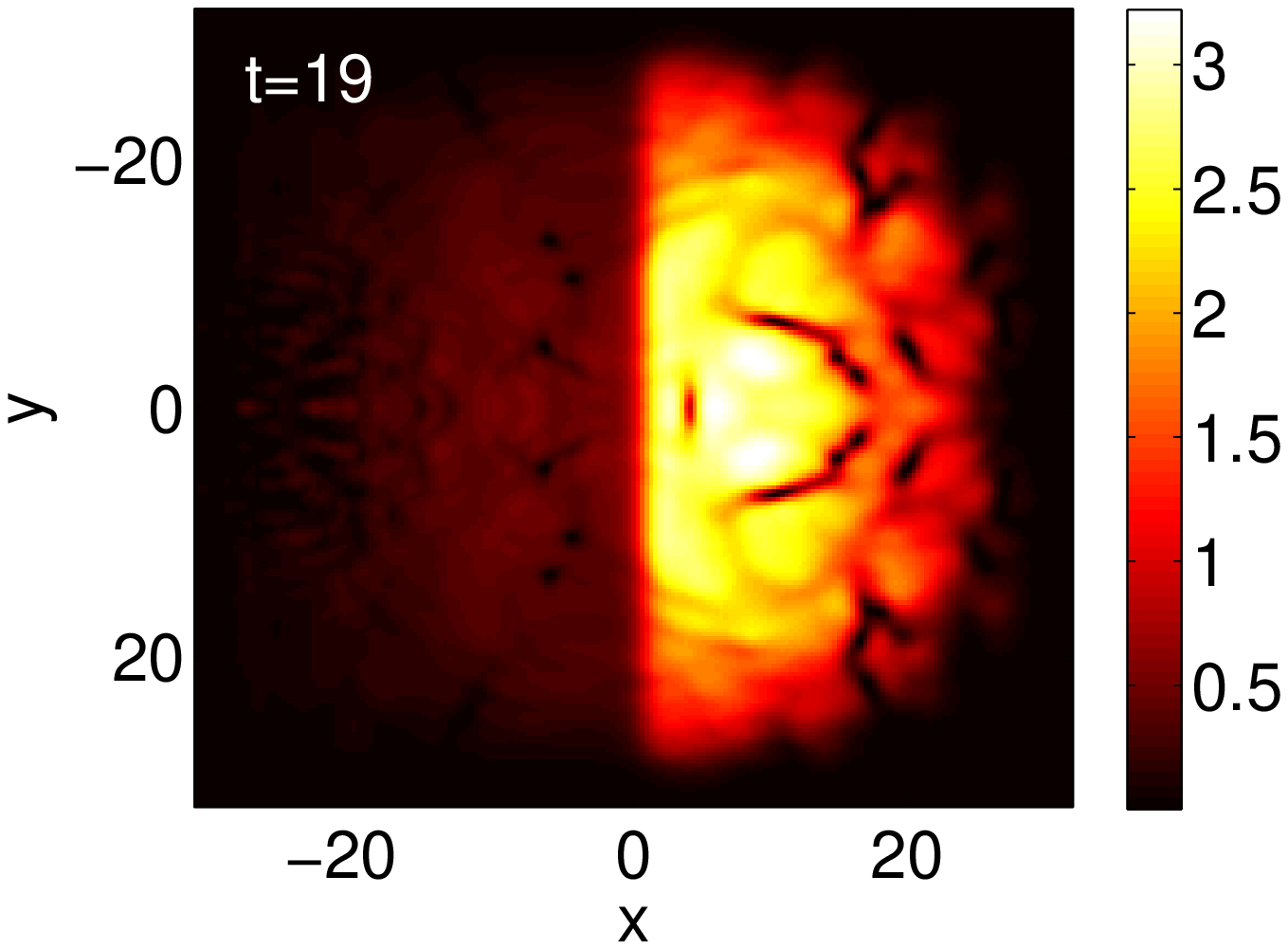}\newline
\includegraphics[width=.22\textwidth]{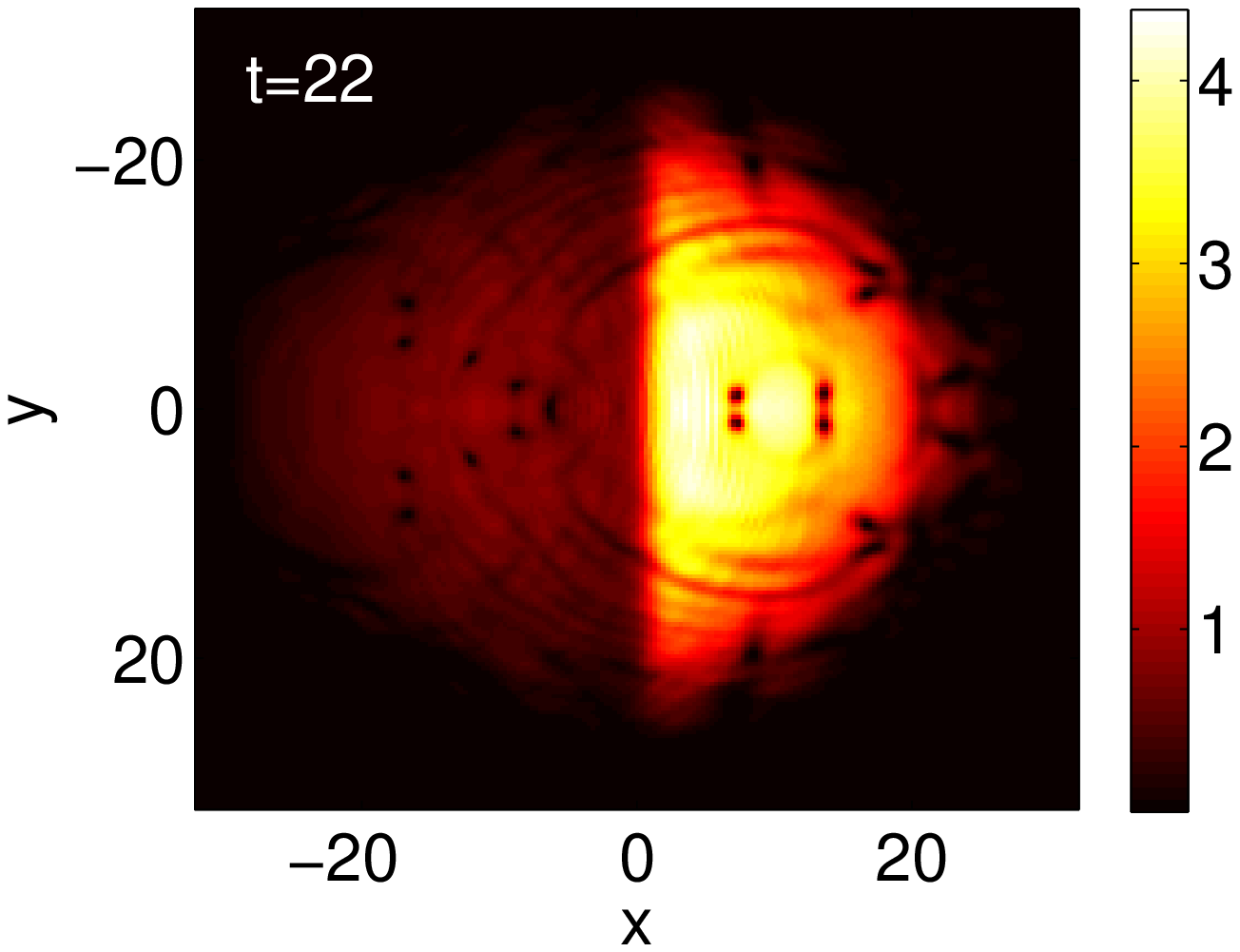}
\includegraphics[width=.22\textwidth]{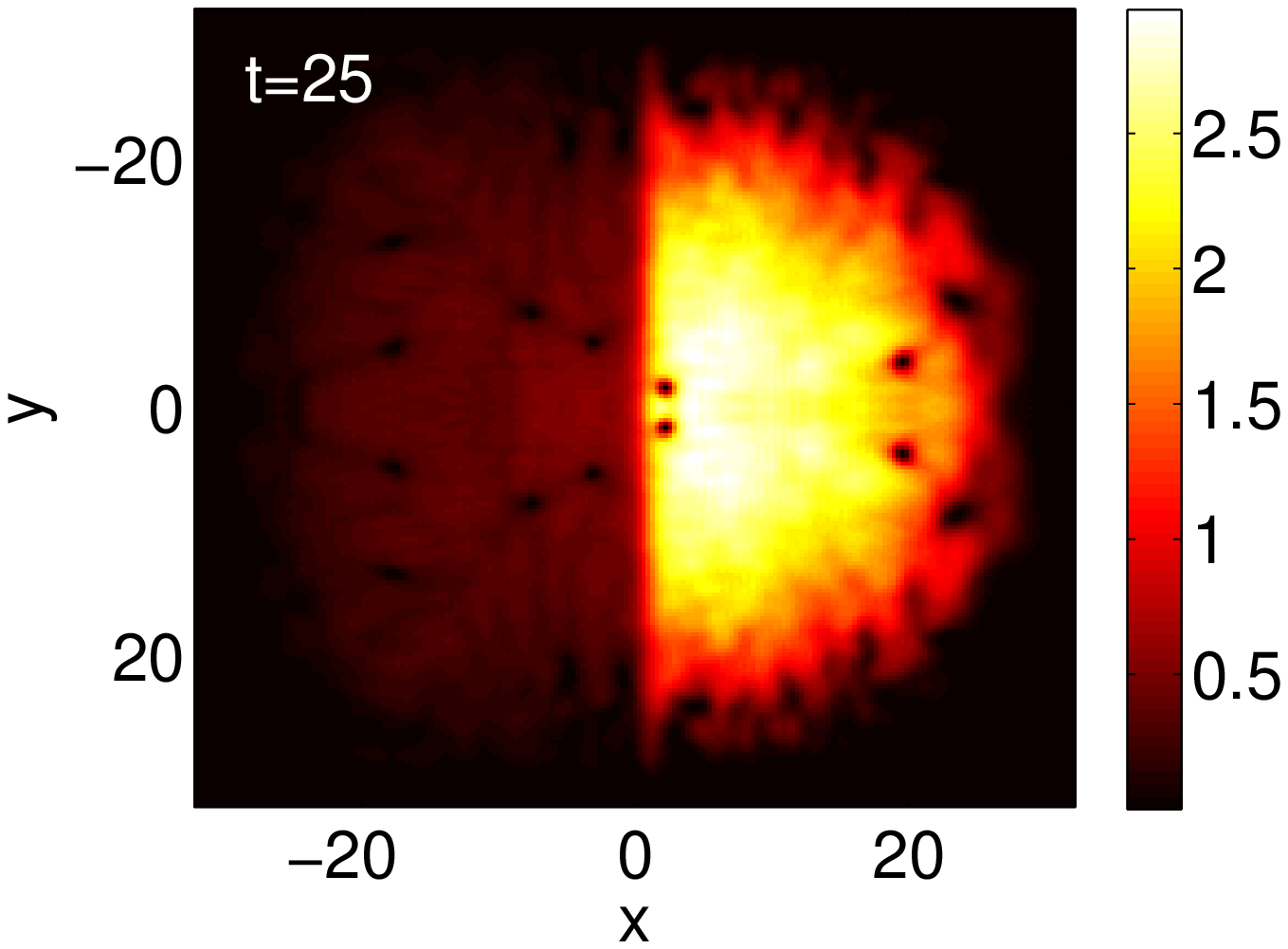}
\includegraphics[width=.22\textwidth]{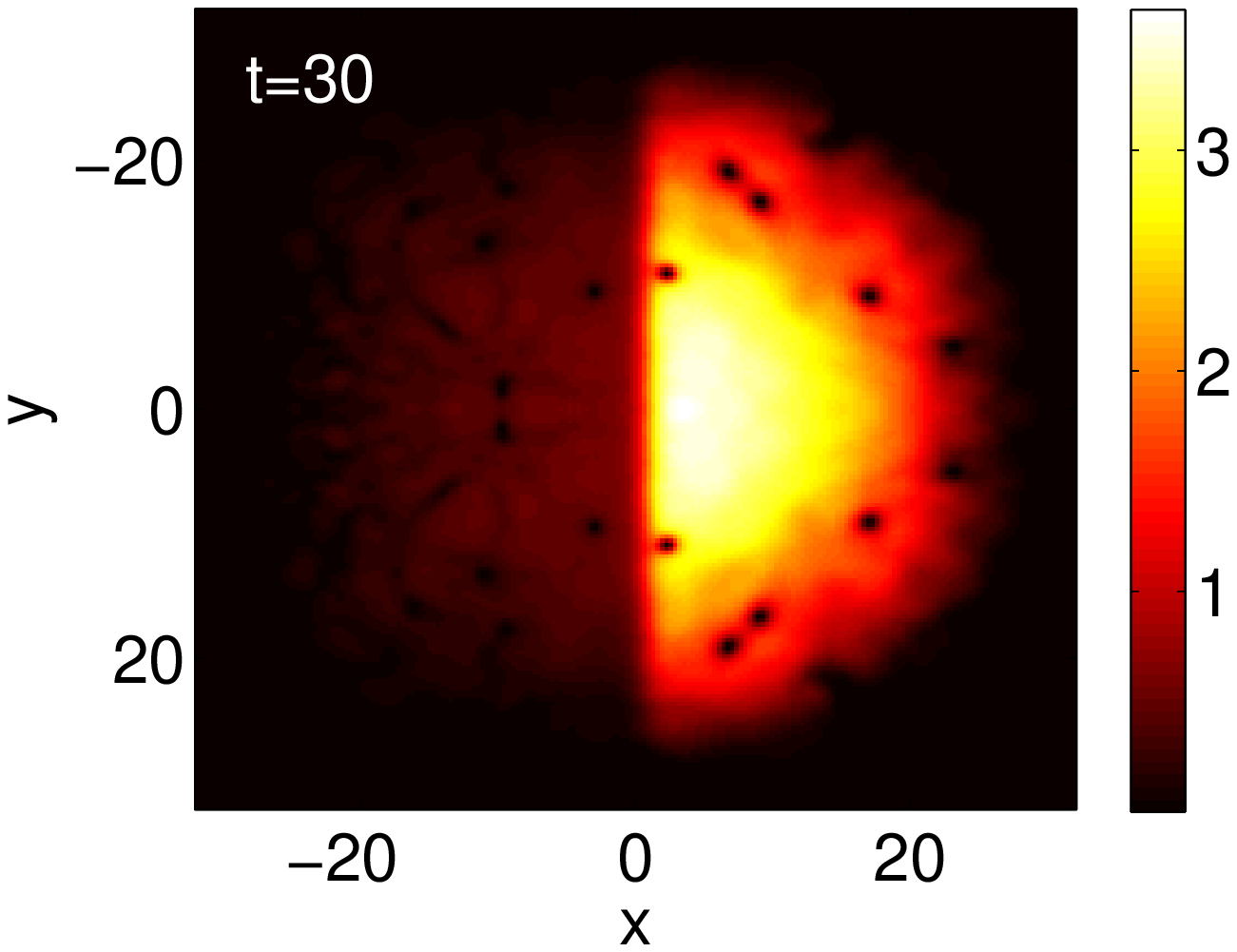}
\includegraphics[width=.22\textwidth]{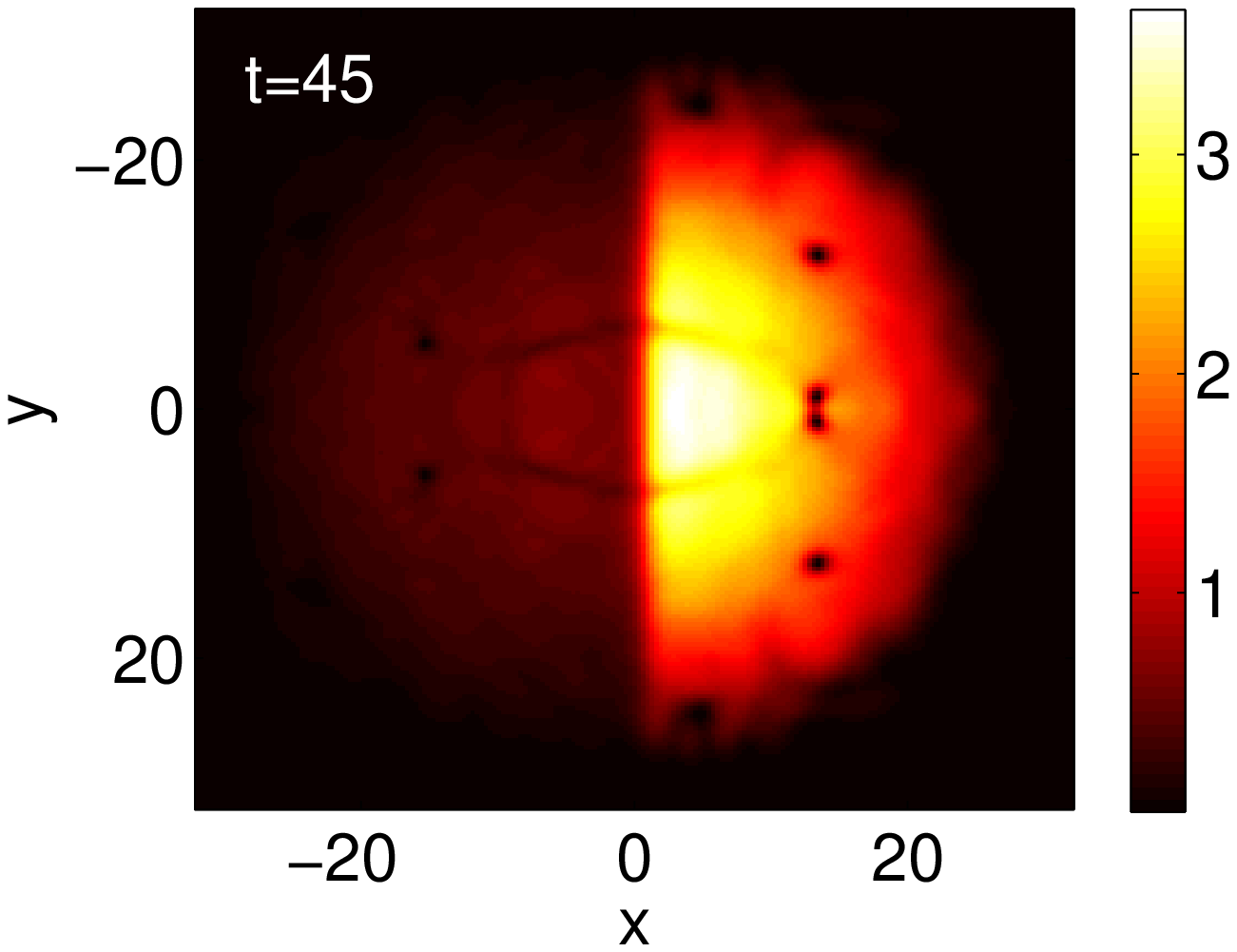}\newline
\caption{(Color online)
Evolution of the 2D condensate density with $\Omega=0.1$ and $\varepsilon=5.0$}
\label{fig6}
\end{figure}

For a given $\Omega$, more dark solitons are emitted when $\varepsilon$ increases.
As expected, and similarly to the above case, the solitons decay again to vortices due to
the onset of the snaking instability: in fact, this happens shortly after their emergence (and their subsequent forward motion for short time). Here, it should be mentioned, however, that
the evolution of the vortices differs in each different case. To be more specific, we provide
two more examples, one with three solitons and the other with two solitons, 
shown in the middle and bottom panels of Fig.~\ref{fig4} 
(parameter values are 
%$\Omega=0.1$ and $0.15$, $\varepsilon=6$ and $7$, respectively). 
$\Omega=0.1$ and $0.12$, $\varepsilon=5$ and $7.5$, respectively).
A series of snapshots of the case corresponding to the middle panels of Fig.~\ref{fig4} is provided in Fig. \ref{fig6}, which demonstrates the emission of the 
two 
dark solitons, together with the evolution of vortices.

\section{Conclusions}
\label{conc}

In this work, we have proposed and analyzed an experimentally relevant technique for the creation of dark solitons and vortex structures in Bose-Einstein condensates. Our method, can be briefly described as follows. After creating a condensate, characterized by a specific scattering length, an external magnetic or optical field (with a value relatively close to a Feshbach resonance) is switched-on. If this external field is spatially inhomogeneous, i.e., it has a steep localized spatial gradient on top of a constant value, then the scattering length becomes spatially inhomogeneous too, taking different values to the left and to the right of the trap center. We have used a mean-field model in $(1+1)$- and $(2+1)$-dimensions, describing cigar-shaped and disk-shaped condensates, respectively, in order to investigate the effect of such a collisional inhomogeneity.
%; the considered model was assumed to have a generalized (non-cubic) nonlinearity, thus taking into account effects due to the deviation of the one- or two-dimensionality. 

Using direct numerical simulations, we have shown that if the change of the value of the scattering length
exceeds a certain threshold, matter-wave dark solitons are spontaneously generated in the system by means of a continuous process. We have found how this threshold for soliton generation depends on other parameter values, e.g., in the quasi-1D setting it becomes smaller when the chemical potential (i.e., the number of atoms) is increased. We have also shown that rectilinear dark solitons can also be spontaneously formed in a disk-shaped condensate. Nevertheless, due to the higher dimensionality of this quasi-2D condensate, the dark solitons are subject to the snaking instability. Thus, shortly after their formation, they decay into vortex pairs. In certain cases, we have also observed the formation of interesting transient structures, in the form   of closed-loop patterns, reminiscent of ring dark solitons.

At this point, it is relevant to mention that, very recently, a detailed experimental control of the inter-atomic interactions in a Bose-Einstein condensate via optical Feshbach resonances was reported \cite{expgofx}. We believe that, as further experiments in the same direction are expected to appear, our results are
likely to be useful as a systematic means of producing fundamental nonlinear
excitations in such collisionally inhomogeneous atomic systems.

\section*{Acknowledgments} The work of D.J.F. was partially supported by the Special Account for Research Grants of the University of Athens.
P.G.K. gratefully acknowledges support from NSF-DMS-0349023 (CAREER),
NSF-DMS-0806762 and the Alexander von Humboldt Foundation. 
P.G.K. also gratefully acknowledges inspiring discussions with
V.M. P{\'e}rez-Garc{\'i}a which initiated his interest in this 
theme.

\end{document}